# Towards Understanding the Structure, Dynamics and Bio-activity of Diabetic Drug Metformin

Sayantan Mondal, Rudra N. Samajdar, Saumyak Mukherjee, Aninda J. Bhattacharyya and Biman Bagchi*

Solid State and Structural Chemistry Unit
Indian Institute of Science, Bangalore-560 012, India
(*email-id: profbiman@gmail.com)

*Abstract*

Small molecules are often found to exhibit extraordinarily diverse biological activities. Metformin is one of them. It is widely used as anti-diabetic drug for type-two diabetes. In addition to that, metformin hydrochloride shows anti-tumour activities and increases the survival rate of patients suffering from certain types of cancer namely colorectal, breast, pancreas and prostate cancer. However, theoretical studies of structure and dynamics of metformin have not yet been fully explored. In this work, we investigate the characteristic structural and dynamical features of three mono-protonated forms of metformin hydrochloride with the help of experiments, quantum chemical calculations and atomistic molecular dynamics simulations. We validate our force field by comparing simulation results to that of the experimental findings. Nevertheless, we discover that the non-planar tautomeric form is the most stable. Metformin forms strong hydrogen bonds with surrounding water molecules and its solvation dynamics show unique features. Because of an extended positive charge distribution, metformin possesses features of being a permanent cationic partner toward several targets. We study its interaction and binding ability with DNA using UV spectroscopy, circular dichroism, fluorimetry and metadynamics simulation. We find a non-intercalating mode of interaction. Metformin feasibly forms a minor/major groove-bound state within a few tens of nanoseconds, preferably with AT rich domains. A significant decrease in the free-energy of binding is observed when it binds to a minor groove of DNA.

- **INTRODUCTION**

In medical applications, small drug molecules display amazing conspicuous remedial actions, often life-saving. In most cases, the reasons for these striking biological activities are not understood at a molecular level. Sometimes the cause of the remedial action lies at the receptors, sometimes at proteins and sometimes with drug-DNA interactions[1]. A detailed molecular level understanding of the action can pave the way for development of better drugs. Sometimes the drug molecules show unique and unusual characteristics. Thus, the physical chemistry of these drugs can be of great interest as a novel physical chemistry problem.

The small molecule Metformin (N,N-dimethylbiguanidine), in its protonated form as Metformin Hydrochloride, is used worldwide as a front line anti-diabetic drug for type-two diabetes.[2-3] Hyperinsulinemia and insulin resistance by the cells are the primary reasons for this disease. For diabetic patients, metformin hydrochloride (hereafter referred to as MET) works by improving the insulin sensitivity of the cells. As a result, the cell receptors can recognize insulin molecules and insulin gets diluted in the blood stream. Apart from this well-studied part, MET also plays quite diverse, lesser known, role. Recent studies report a possible connection between increased risk of certain types of cancers and type-two diabetes[4]. MET activates adenosine monophosphate activated protein kinase (AMPK) that results in a decreased rate of tumor growth.[5] It also targets certain types of cancer cells and oncogenes both *in vivo* and *in vitro*.[6-8] Clinical oncologists often use MET as a potential anti-tumor agent simultaneously with chemotherapy on cancer patients.[9-11]

MET is available as an over-the-counter drug in an affordable rate than other medicines of similar category. The dose may vary from 50mg to 2000mg per day[12]. Due to the high doses, in some parts of the world, metformin is found in water bodies and sewages which causes water pollution and hampers marine bio-diversity[13]. Whether there are a number of facts and information available in terms of medical knowledge and mere speculations, a detailed theoretical investigation at the molecular level is clearly lacking. This should be the first step towards understanding the mechanism and drug action of not only MET but also many other biologically important small molecules, which is, as of now, difficult and only partially understood.[14-15]

There exist thermodynamic and kinetic studies of DNA intercalation of several small molecules like daunomycin, *cis*-platin, proflavine etc.[16-20] However, relatively lesser number of theoretical studies are present.[21-25] The general pathway of drug intercalation can be thought of as a combination of diffusion, outside groove binding and finally intercalation. However there are other kinetic models as well. Some molecules, for example, Hoechst, DAPI, methyl red etc. are found not to intercalate at all. These form a groove bound state with DNA but may possess anti-tumor and antimicrobial properties.[26] The interaction of small molecules with DNA is also associated with DNA damage.[27]

Although studies on the binding/ intercalation/ interaction of small molecules to protein/DNA has remained in the frontline for many decades, especially in disciplines like molecular biology and biophysical chemistry, our understanding of such processes has been slow to develop, because of the immense complexity of the systems involved. The delocalized positive charge on metformin hydrochloride makes it a potential ligand for DNA binding. Although there are no solved crystallographic or microscopy data for DNA-Metformin complex, the pre-intercalative outside bound state draws avid attention. However, the crystal structure of metastable metformin which is important for any medicine that are distributed in the solid phase, is recently solved[28]. Experimental studies of its interaction with biomolecules are also emerging.[18, 29]

In this paper, for the first time, we present a detailed molecular level study of metformin hydrochloride and its interaction with DNA with the help of computer simulations, quantum chemical calculations and spectroscopy. In the present work, we first characterize the molecule MET by investigating some of its structural and dynamical properties in aqueous solution. We perform quantum chemical calculations and molecular dynamics simulations to describe the molecule along with its different protonation states. We find that the non-planar protonated form exhibits maximum stability. We use an AMBER based force-field description. The validity of thus obtained force field is checked with respect to experiments. We then perform free energy calculations to understand the pathway of DNA binding with the help of metadynamics based sampling techniques.

- **COMPUTATIONAL METHODS**

**Quantum chemical calculations and force field.** In aqueous solution and at physiological pH (~6.9-7.4), metformin remains in the mono-protonated state. There are two possible protonation sites (**Figure 1b**, **1c**) and these two protonated forms can tautomerise to produce another form (**Figure 1d**). We arrive at the energy minimized structure for each molecule using density functional theory (DFT) calculation with B3LYP functional and 6-31G + (d, p) basis set. Water is considered as an implicit solvent to provide a dielectric continuum during the optimisation. The optimized structures are further optimized using second order Møller-Plesset (MP2) perturbation theory method with the combination of augmented correlation consistent double zeta (aug-cc-pVDZ) standard basis set in order to obtain the final geometry and relative energies. We use this information to build the topology for simulation. In order to obtain the UV-VIS spectra, time dependent density functional theory (TDDFT) is applied for 10 states. All quantum calculations are performed with Gaussian 09 package[30].

**System and simulation details.** Based on the relative stability and presumable co-existence, we choose three structures for molecular dynamics simulations (MET1, MET2 and MET3) as described in **Figure 1**. The geometry information (bonds, angles and dihedrals) are obtained from the respective optimised structures. Charges are calculated by RESP method that considers equivalence of atomic positions[31]. Other force field parameters (for example, LJ parameters, force constants etc.) are obtained by comparing the atom types with generalized amber force field (GAFF)[32] atom types and using antechamber package (force field and topology parameters are provided in **Supporting Information**).

We perform atomistic molecular dynamics simulations using GROMACS[33] (v 5.0.7). Three different systems are set up for three different molecules as mentioned above. For each system, a single metformin molecule is solvated in ~500 TIP3P water molecules inside a cubic cell to comply with the experimentally studied concentration. In order to neutralize the charges chloride ions are added. We study the system size dependence of diffusion in 3 more systems that are generated in the same concentration (with 1440, 2880 and 4320 water molecules).

For free energy studies of DNA binding, we prepare two model B-DNA dodecamer using nucleic acid builder (NAB) module implemented in AMBER[34]. The sequences are d(GCGCTCGAGCGC)$_2$ and d(ATATCTAGATAT)$_2$. As there are no solved crystal structures or microscopy image of DNA-MET complex, we prepare complex with the help of Maestro code in Schrödinger package.[35] Here we show the free energy calculation for only DNA-MET3 complex as it is the most stable one in the physiological conditions. The DNA molecules are described using Amber99sb-ildn force field[36] which is compatible with small molecules described by GAFF[32]. The prepared DNA-MET3 complex is solvated into 21010 TIP3P water molecules.

Charge neutralization is achieved by adding 22 $Na^+$ and 1 $Cl^-$ ions. Periodic boundary conditions are imposed in three dimensions.

The systems (described above) are energy minimized using steepest descent method followed by conjugate gradient method. Simulated annealing is applied starting from 300K to 330K and again reducing the temperature slowly to 300K. We equilibrated the systems first by restraining the positions of the solutes for 2ns and then by removing the restrain for another 5ns in NPT ensemble (300K, 1bar). Then the equilibrated systems are subjected to 5ns NVT (300 K) simulations in order to achieve the desired equilibrium states. For MET-Water systems the final production runs are carried out for 10 ns. Metadynamics simulation details and description of order parameters are elaborated in Sec. G under 'results and discussions'.

Modified Berendsen thermostat and Parrinello-Rahman barostat are used to keep the average temperature and pressure constant respectively. The equations of motion are integrated using a MD time-step of 2 fs. The cut-off radius for neighbor searching and non-bonded interactions is taken to be 10 Å. All bonds are constrained using LINCS algorithm. PME is used for calculation of electrostatic interactions with FFT grid spacing of 1.6 Å.

- **RESULTS AND DISCUSSION**

(A) Structure and Protonation States

Metformin has more than one protonation sites that are crucial for determining its structure in the aqueous medium. $pK_1$ and $pK_2$ of neutral metformin molecule in gas phase is measured to be 3.1 and 13.8 respectively. Metformin is neutral when the pH is greater than $pK_2$. When the pH is in between $pK_1$ and $pK_2$ (that is, physiological pH), metformin captures a single proton and remains positively charged. In conditions pH<pK$_1$, it goes to a di-protonated form. We study only the mono-protonated forms as these are biologically important.

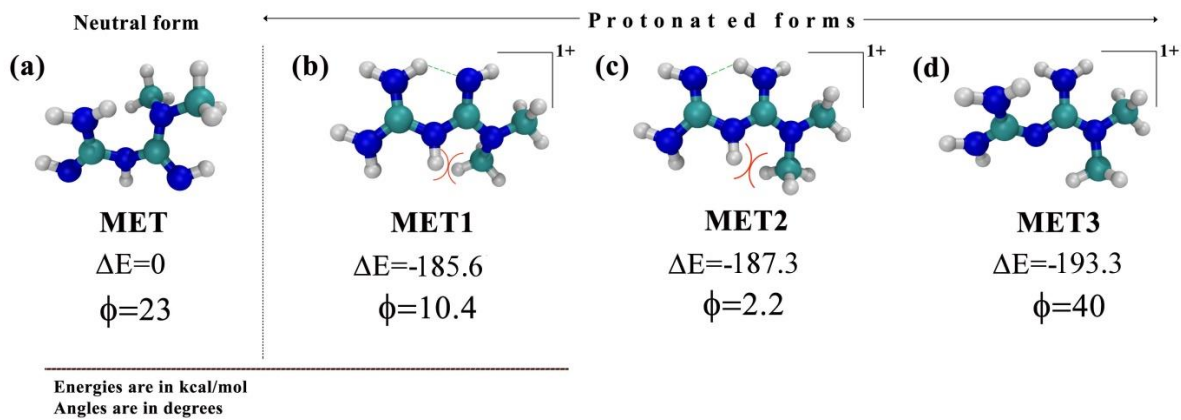

Figure 1. Various protonation states of metformin, their geometry and relative stability as obtained from MP2 level optimization and energy minimization in water as an implicit solvent. Phi (∅) denotes the angle between the planes of two guanidine groups. It is seen that planar forms (b) and (c) are of comparable stability but the tautomeric non-planar protonated form (d) is of the lowest energy. The stability of the non-planar form is because of the release of the steric strain with a hydrogen and methyl groups as indicated in the figure in red.

The neutral form in water is not planar. We choose the energy of the neutral form along with an isolated proton in dielectric continuum as the reference frame for energy. With respect to this, MET1, MET2 and MET3 possess stabilization energies of ~185.6 kcal/mol, ~187.3 kcal/mol and ~193.3 kcal/mol (**Figure 1**). We have performed another set of optimization and energy calculation by using the same level of theory but in the presence of explicit water molecules in the first hydration shell. As before, we set the reference to the energy of the neutral form and an isolated proton. With respect to this reference, MET1, MET2 and MET3 possess relative stabilization energy of ~150.7 kcal/mol, ~155.2 kcal/mol and ~157.7 kcal/mol respectively. That is, all the protonated forms are substantially more stable compared to the neutral one; as shown by our quantum calculations. Addition of a proton induces planarity to the system. But, surprisingly the most stable form, that is the tautomeric form MET3, is non-planar.

We can rationalize this extra stability of MET3 in the following fashion. We note that the conjugation is preserved in the non-planar form. We note that the two guanidine subunits retain their charge delocalization. In MET1 and MET2 there is one stabilizing hydrogen bond interaction (shown by green dashed lines in **Figure 1**) and another destabilizing steric hindrance between one methyl group and an hydrogen atom in the middle (shown by red color in **Figure 1**). However, MET3, because of its puckered structure, loses the stabilization through hydrogen bond but at the same time, overcomes the steric strain. We speculate the steric effect predominates and makes MET3 the most stable mono-protonated form.

## (B) Choice of Protonation States and Validation of Force-field

In order to identify the predominant protonated form in nature and validate the force field which we use in subsequent atomistic molecular dynamics simulations, we carry out experiments to determine a few basic properties like diffusion coefficient using diffusion ordered NMR spectroscopy (DOSY). The spectral characterization is also done with respect to UV-VIS, IR and NMR to get structural information. Experimental results are compared to those of theoretical findings.

**Crystallization and purity of the drug**. We isolate crystals of pure metformin hydrochloride from Glycomet IP 500 mg, a commercially available prescription drug containing metformin, manufactured by US Vitamins Limited. Two 500mg tablets are ground thoroughly using mortar – pestle and stirred in water at 70° C for seventy two (72) hours. The resultant suspension is filtered and the clear filtrate thus obtained is cooled slowly over a couple of days to yield needle shaped white crystals of pure metformin. We characterize the purity of the sample using $^1$H

NMR. A few milligrams of the crystal is dissolved in D$_2$O (99.9% D, Sigma Aldrich) and NMR data is recorded in a Bruker 400 MHz machine. Bruker Top Spin processing software is used for the data analysis. The observed NMR spectrum shows only one compound peak at 3.03 δ ppm (**Supporting Figure S1**) corresponding to the methyl protons of metformin structure, thus confirming the purity of the crystals obtained (indexed to The Human Metabolome Database entry HMDB0001921)[37]. The crystals are found to melt between 232 – 234°C. Melting of different polymorphs of metformin hydrochloride has been reported in literature to occur within the temperature range 224° C to 232° C.[38-39]

UV-VIS spectroscopy. We perform UV Visible absorption experiments using a 50 μM solution of Metformin in water in 10 mm path length quartz cuvette, on a Perkin Elmer Lambda 750 spectrophotometer. The wavelength window used is 200 – 800 nm. The UV-Vis absorption spectrum of Metformin in water shows a strong absorption centered around 230 nm which arises due to n→π* transition characteristic of the imine moiety[40]. Variation of concentration from 15 μM to 150 μM shows no significant changes in the absorption spectrum, apart from the expected linear increase in absorption of the 230 nm peak (**Supporting Figure S2**). A resultant Lambert – Beer analysis is used to estimate the molar extinction coefficient as 10240 (±230) M$^{-1}$cm$^{-1}$. TDDFT calculations provide the probable UV - visible spectra for different protonated species (**Figure 2**).

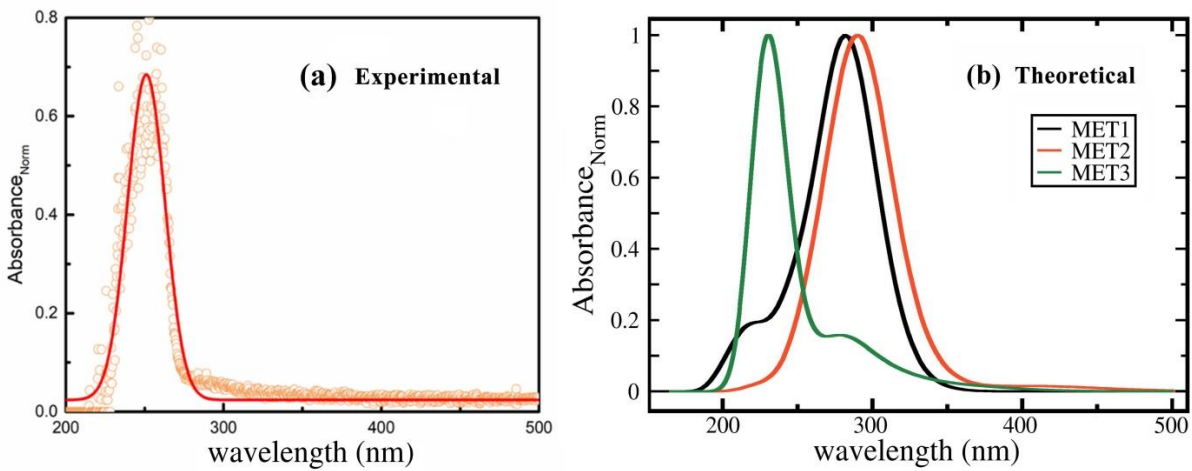

**Figure 2.** Characterization of the protonate state of metformin is demonstrated by combining UV – visible absorption spectrum of metformin obtained from experiment (A: orange circles represent the data points, the orange trace represent the Gaussian fit) and by TDDFT calculations for three different protonated species (B). In the experimental spectrum the absorption peak is at ~230 nm. The theoretically calculated spectrum for MET3 (~235 nm) is closest to the experimentally observed absorption spectrum of metformin hydrochloride in solution which indicates the predominance of the form MET3 in the experimental sample.

Out of the three probable forms, the one that is found to be closest to experimental data corresponds to the tautomeric form MET3, which, we conclude, is the predominant form in aqueous solution. The theoretically calculated absorbance peak for MET3 cantered at 235 nm is in good agreement with the experimentally observed absorption spectrum at 230 nm. So we draw most of our conclusions by studying this tautomeric form alone. This form shall be used later when we study its interaction with DNA.

**Diffusion Coefficient Measurements.** $^1$H Diffusion Ordered Spectroscopy (DOSY) is performed on a Bruker 400 MHz spectrometer using the NMR sample prepared as discussed earlier. Bruker TopSpin processing software is used for the data analysis. The diffusion coefficient corresponding to the metformin molecule in $D_2O$ is found to be $0.98 \times 10^{-5}$ cm$^2$/s from DOSY spectrum. This value of D is used to estimate the hydrodynamic radius ($r_H$) of metformin using the Stokes-Einstein equation shown below.

$$D = \frac{k_B T}{6\pi\eta r_H} \quad (1)$$

Here viscosity coefficient ($\eta$) is $1.2503 \times 10^{-3}$ Nm$^{-2}$s for $D_2O$[41] and the calculated $r_H$ is 1.76 Å. The diffusion coefficient obtained from DOSY is $0.98 \times 10^{-5}$ cm$^2$/s, which is again closer to the calculated diffusion coefficient in water at similar conditions from MD simulations.

In our computer simulation studies, we use mean squared displacement (MSD) to find diffusion constant from the slope of MSD vs time graph with the help of the following Einstein equation.

$$<\Delta r(t)^2> = 6Dt \quad (2)$$

The deviation from experiment is small which provides strong support in favor of the force field of metformin (**Figure 3** and **Table 1**). As the calculation of diffusion constant is dependent on the system size we perform four simulations with different system size and report the values for the largest system size (that is 9 MET in 4320 water molecules) in **Table 1.** The dependence of diffusion constant on system sizes is discussed in the ***Supporting Information*** (**Table S1**) section. The trend in the diffusion coefficients can be explained in terms of the average hydrogen bond lifetime of MET with water (**Table 3**). The average H-bond lifetime for MET1, MET2 and MET3 are 5.8ps, 5.1ps and 4.5ps respectively. That indicates a lesser resistance experienced by MET3 which eventually increases its diffusion in liquid water.

However, the diffusion coefficient values do not demonstrate the abundance of MET3 in the experimental sample. The values obtained from simulations of different protonated forms are fairly close to each other. We note that the solvent used in DOSY experiments is heavy water (η~1.25 mPa.s) as opposed to that of the neat water (η~1 mPa.s) used in simulations. As the coefficient of viscosity of $D_2O$ is 25% higher than $H_2O$, it only scales the diffusion coefficient by ~25% as Eq. (1) is quantitatively valid in this range. Hence, the calculated diffusion constants can be compared with the experiments.

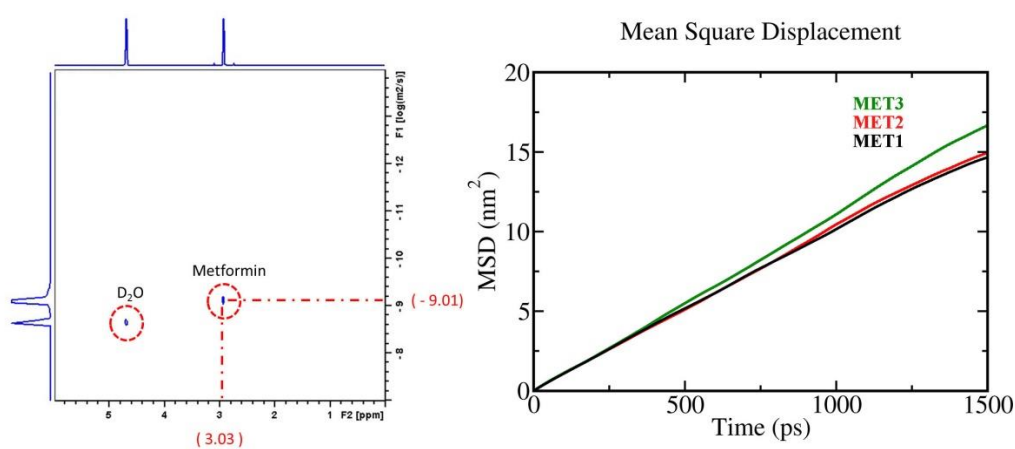

**Figure 3. DOSY spectrum of metformin hydrochloride in $D_2O$ showing the diffusion constant corresponding to metformin molecule and $D_2O$ (left); MD simulation data used for estimation of diffusion coefficient of metformin in its various protonated forms using mean squared displacement versus time plot (right) - MET1 (black), MET2 (red) and MET3 (green).**

**Table 1**. **Diffusion coefficient of metformin as calculated from DOSY experiment and separately from classical molecular dynamics simulation. The results are noted down here for the largest simulation box (that is with 4320 water and 9 metformin molecules). Dependence on system size is given in the supporting information section (Table S1).**

|      | D (experimental) cm$^2$/s | D (simulation, in H$_2$O) cm$^2$/s |
|------|---------------------------|-------------------------------------|
| MET1 | $0.98 \times 10^{-5}$ (in D$_2$O) | $1.13(\pm 0.35) \times 10^{-5}$ |
| MET2 | [that is, $1.23 \times 10^{-5}$ (in H$_2$O) by viscosity scaling according to Eq.1 ] | $1.25(\pm 0.45) \times 10^{-5}$ |
| MET3 |  | $1.47(\pm 0.53) \times 10^{-5}$ |

**Infra-red Spectroscopy.** We use IR spectroscopy to identify the vibrational modes of different bonds in metformin. IR experiments are carried out in ATR mode using pure crystals of the drug on a Perkin Elmer Frontier instrument within the wavenumber window of 400 – 4000 cm$^{-1}$. The IR spectrum of the compound is shown in **Figure 4**. The primary bands seen in IR are indexed to vibrational modes expected in the metformin molecule (**Table 2**) using standard data available in literature[42].

We next compare the experimentally obtained spectrum with that from quantum chemical calculations. **Table 2** contains the detailed description of the absorption bands.

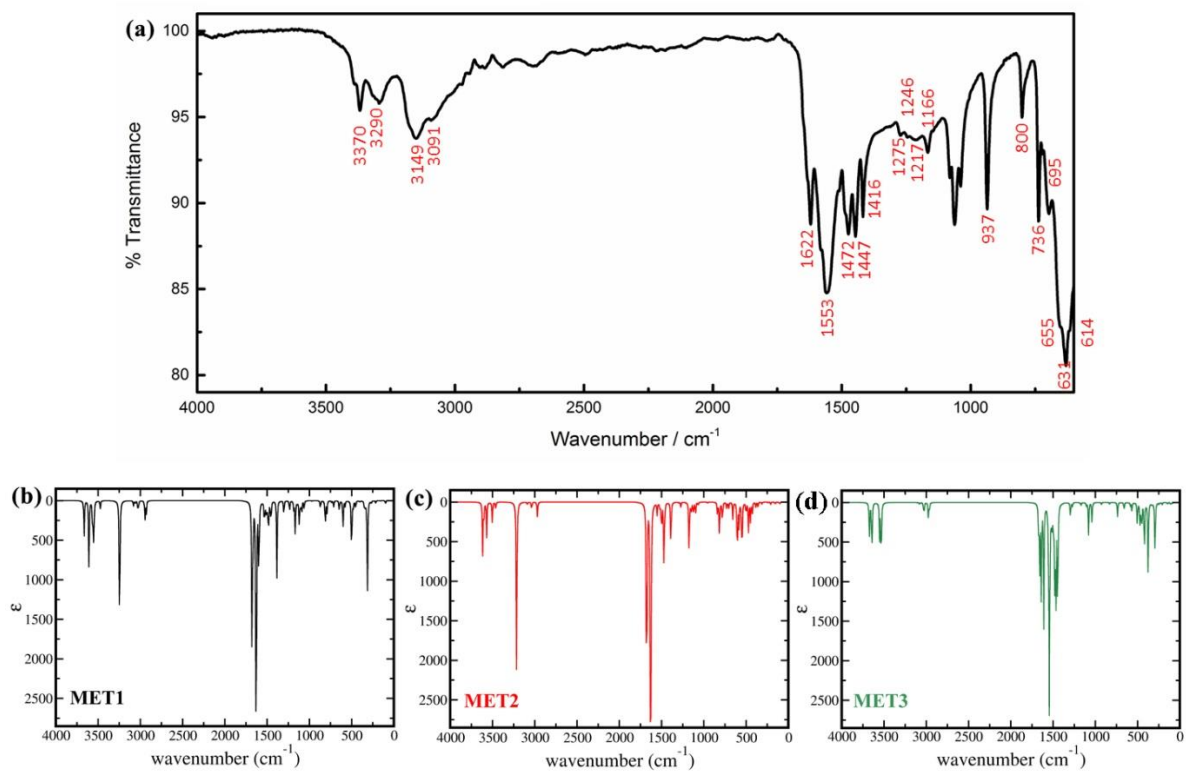

**Figure 4. Characterization of protonation states of metformin with the help of Infrared spectrum (a) experimentally and (b)-(d) with the help of quantum calculations on three different protonation states in implicit water solvent. It is clear that the tautomeric form resembles more with the experimental spectra. This again proves the dominance of the tautomeric form in the experimental sample.**

The broad absorption band at ~1400-1600 cm$^{-1}$ is the characteristic absorption band for N-H bending mode of imine group. This feature is also present in the theoretically calculated spectrums. Additionally there is considerable absorption ~3200 cm$^{-1}$ for MET1 and MET2 forms, as obtained from theory. *This band is absent in MET3 as well as in the experimental spectra. This again proves the dominance of MET3 form among other protonated forms.* In the later section where we discuss the interaction with DNA, only MET3 is considered.

**Table 2. Vibrational modes identified from IR Spectrum of metformin**

| Wave number / cm$^{-1}$ | Vibrational Mode |
|---|---|
| 3370 | C=N-H (Imine) |
| 3290, 3149 | N-H stretching / symmetric |
| 3901 | C-H stretching / asymmetric |
| 1622 | >C=N stretching |
| 1553 | N-H bending vibration / Imine |
| 1472, 1447, 1416 | Asymmetric C-H deformation in N(CH$_3$)$_2$ |
| 1275, 1246, 1217, 1166 | NH$_2$ rocking / twisting |
| 1081, 1062, 1040 | CH$_3$ rocking vibration |
| 937 | Symmetric C-C=N stretch |
| 800, 736, 695, 655 | N-H out of plane bending |

**Melting Studies from Simulation.** Melting studies are yet another way to check the robustness of the force field. The crystal structure of a polymorph of the tautomeric form of metformin is solved by Patrick *et al.*[28] This is again the tautomeric form that we claim is the most abundant. We use the crystal structure to build a monoclinic simulation box (4.077nm ✗ 7.154nm ✗ 3.717nm and α=90°=γ, β=114°) of 450 MET3 molecules (that is 450 metformin-H$^{(+)}$ and 450 Cl$^-$). By the use of simulated annealing technique, we rise the temperature from 0K to 1000K in steps of 20K and equilibrated the system (NPT) until it attains a stable value of density. Each configuration is further equilibrated for 100 ps in NVE ensemble. Hence, there are fifty trajectories that correspond to fifty different temperatures.

Two kinds of analyses are performed to identify the melting point of the crystal. The first one is to investigate the nature of the radial distribution function (RDF) of center of mass of MET. The position and the height of first peak drastically changes near melting point (**Figure 5a** and **5b**). Another way is to monitor the potential energy (Coulomb and LJ) per molecule with temperature. It also shows a break in the slope near the melting point (**Figure 5c**). In **Figure 5d**, we show the variation in density with temperature with a similar break near 568K. All three studies reveal the melting point ~568K whereas the experimentally measured melting point is ~505K. Our force field description overestimates the melting point by ~12% which is usually regarded as a good agreement in the existing literature on force-filed development. The discrepancy may be partly because of the way the force field is developed. It is not suitable to describe the molecule in its solid state. Despite it predicts the melting point only 12% higher.

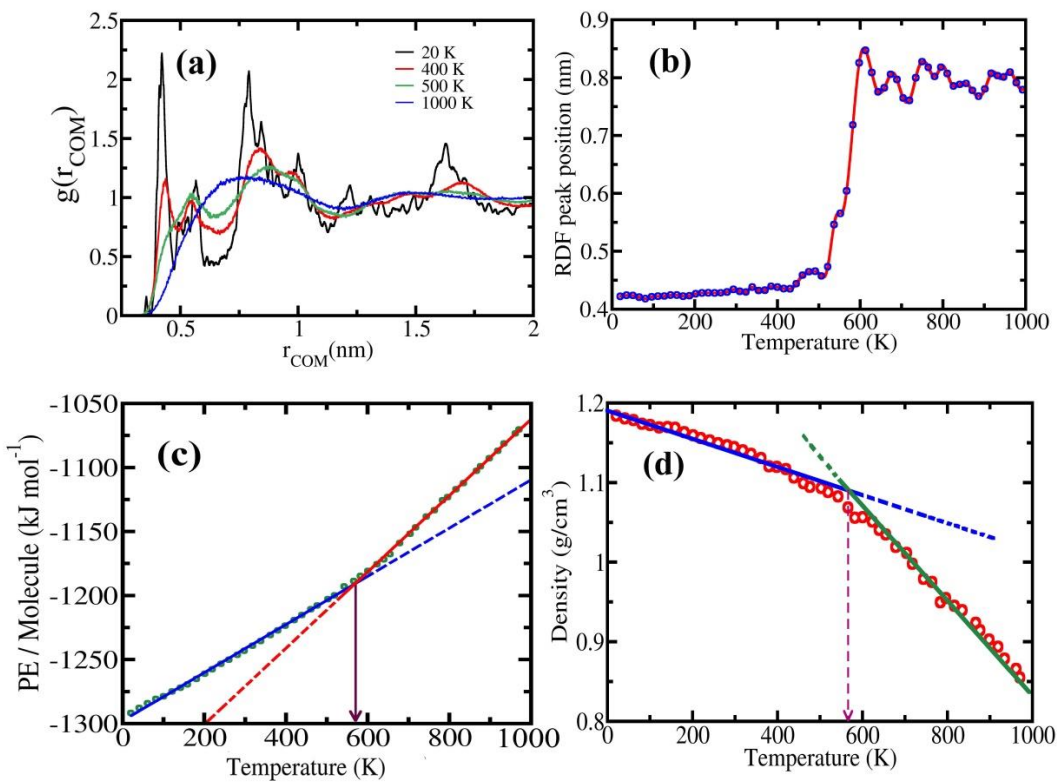

**Figure 5. Broad agreement of the melting point temperature between simulation results and experimental studies of MET crystal as reported by Patrick et al[28]. (a) Exemplary plots of radial distribution function (RDF) of centre of mass of metformin molecules in four different temperatures. The nature of the peaks and their positions dramatically changes when the crystal melts. (b) First peak position versus temperature plot. This shows a discrete jump around 568K. This can be considered as the melting temperature. (c) Potential energy per molecule against temperature also shows a break in the slope near 568K that also suggests melting. (d) Density against temperature plot shows a break in the slope near 568K.**

**Preservation of the Geometry of Metformin.** In order to check the ability to retain the conformation we perform ~100ns well-tempered metadynamics simulation on MET-TIP3P (1:500 molecules) water system. We consider two torsion angles as order parameters with Gaussian hill height of 0.5 kJ/mol and width of 0.05 radians (for dihedrals). It is similar to the Ramachandran's plot in case of peptides. These two dihedral angles are defined in **Figure 6**. It is clear that MET1 and MET2 have the lowest free energy when they are planar. On the other hand some amount of puckering is associated with the free energy minima of MET3. The second patch in the free energy surface of MET3 arises because of another similar conformation when $\phi_1$ rotates by 180°. However up to our several ns during unbiased MD simulation, because of the high energy barrier of escape (greater than 40 kJ/mol), the molecules are able to retain its conformation which corresponds to the deepest free energy minima.

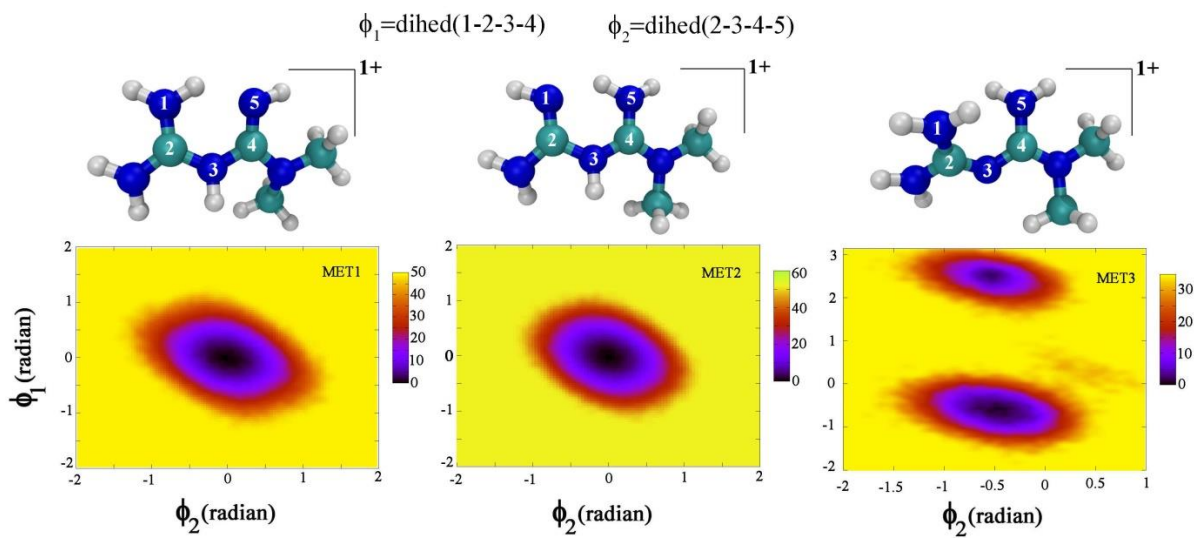

**Figure 6.** Free energy (in kJ/mol unit) contours with respect to two dihedral angles ($\phi_1$ and $\phi_2$) of the three different mono-protonated forms of metformin hydrochloride. Because of the deep minima in free energy all three of them preserve the planar/non-planar conformation to a great extent under our force field description. In the timescale of performed MD simulations they are observed to evolve only in the deepest minima.

## (C) Energy Barrier of Proton Transfer

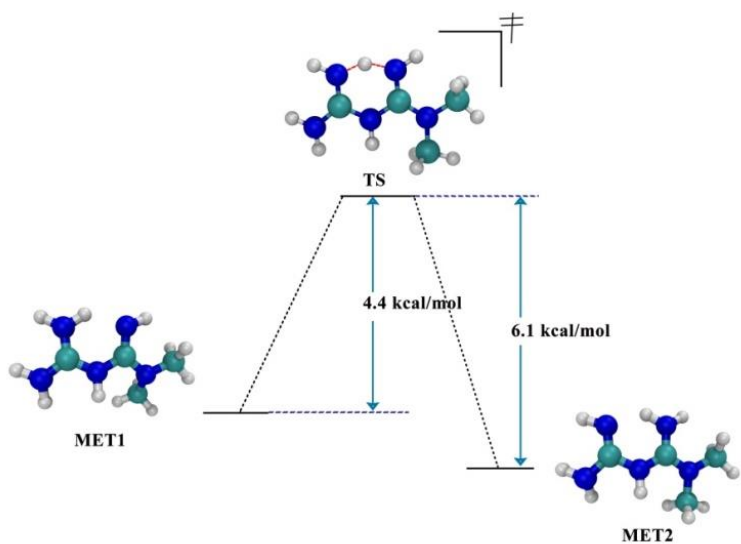

**Figure 7.** Schematic representation of proton transfer reaction from one to another planar mono protonated form. The barrier crossing energy is low (~4-6 kCal/mol) which allows a possible co-existence of these two species in the solution.

In order to find out the possibility of co-existence among the various protonated forms we calculate the energy barrier of proton transfer between MET1 and MET2. The transition state (TS) is optimized by using the same level of theory (optimized to Berny TS) as the stable structures and traced back to the reactant/products by the use of internal reaction coordinate (IRC) scan. We find that the energy barrier for MET1 → MET2 is ~4.4 kcal/mol and for that of MET2 → MET1 is ~6.1 kcal/mol (**Figure 7**). The energy barrier can be crossed in room temperature and allows us to speculate a possible co-existence between the two forms of metformin hydrochloride in the aqueous solution.

## (D) Water Molecules around Metformin

**Radial distribution function.** The partial pair correlation functions [$g_{\alpha\beta}(r)$], where α stands for either N or H of metformin and β stands for the oxygen atom of water molecules, are calculated in order to realise the local structure of water molecules around metformin hydrochloride. $g_{\alpha\beta}(r)$ is calculated from the classical MD trajectories. We match this result to the ab-initio calculations. We put explicit water molecules around metformin (**Figure 8**) and optimized the whole system by using MP2/aug-cc-pVDZ. We find that the water molecules are stabilized at a distance of ~3Å from the N-atoms and ~2Å from the H atoms. This provides some rationalisation in the calculated g(r) from MD simulations (**Figure 9**). Also that the tautomeric form is non-planar and the most stable in explicit water solvent.

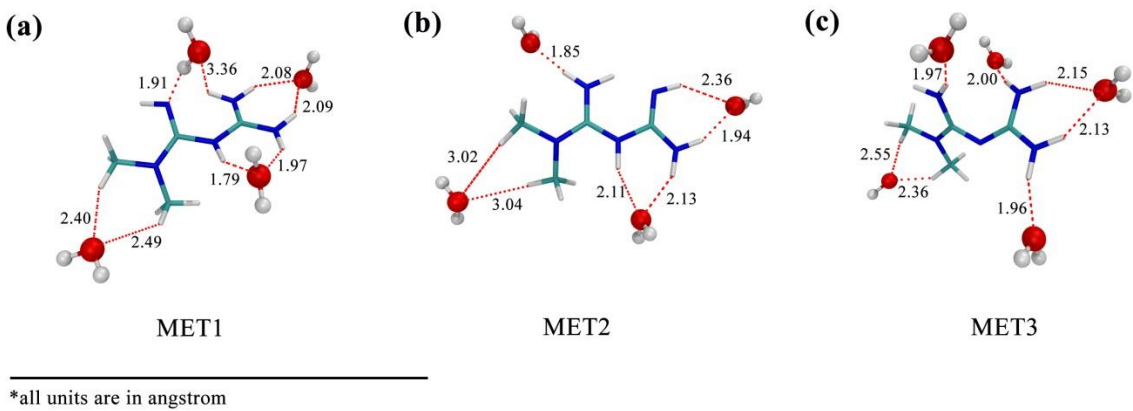

**Figure 8. Representation of the optimised structures of various protonated forms of metformin hydrochloride in explicit water molecules (five) as a solvent. The water molecules are stabilised at ~2-3 Å. We observe that the tautomeric form is the most stable one here as well with a non-planar geometry.**

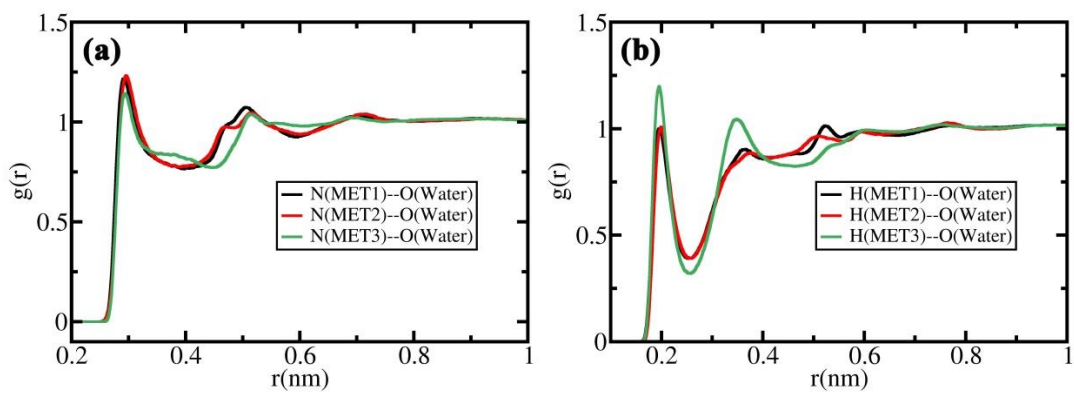

**Figure 9. Radial distribution functions between certain atoms of metformin and oxygen atoms of water. (a) Pair correlation function between nitrogen atoms of metformin and oxygen of water. This shows two peaks ~3Å and 5Å. (b) pair correlation function between acidic hydrogen atoms of metformin and oxygen of water. This shows two peaks ~2Å and 3.5Å.**

**Hydrogen bond dynamics.** Because of the presence of six nitrogenous hydrogen atoms there can be extensive hydrogen bonding between MET and water molecules. The hydrogen bond lifetime and strength are reflected in self-diffusion, orientational mobility and solvation (as discussed later). We employ a technique that was developed by Chandler *et al.* using of Heaviside step function ($H$ or $h$) formalism[43]. We use the usual geometric criteria as the definition of hydrogen bond. $h(t)$ adapts a value equal to '1' if the bond is present and '0' otherwise. On the other hand, $H(t)$ is '1' if the bond is present and '0' for the rest of the trajectory if the bond breaks only once. We can define two types of time correlation functions using these definitions namely continuous hydrogen bond correlation, $S_{HB}(t)$ and intermittent hydrogen bond correlation, $C_{HB}(t)$ as defined in **Eq (3)**.

$$C_{HB}(t) = \frac{\langle h(0)h(t) \rangle}{\langle h \rangle}; \; S_{HB}(t) = \frac{\langle h(0)H(t) \rangle}{\langle h \rangle} \quad (3)$$

Surprisingly the three protonated forms of metformin hydrochloride show quite distinct hydrogen bond dynamics and lifetime (**Figure 10** and **Table 3**). Moreover, the trend is completely opposite in $C_{HB}(t)$ and $S_{HB}(t)$. It happens because water molecules can come near MET and goes away in the next step. But it can again come back and this cycle goes on for a few ps. In that situation $S_{HB}(t)$ picks up '0' forever but $C_{HB}(t)$ oscillates between '0' and '1'.

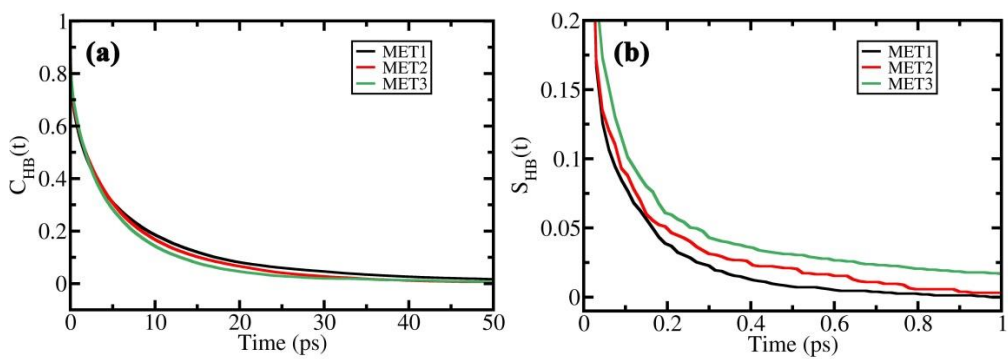

**Figure 10**. **Hydrogen bond dynamics between metformin and water molecules. (a) Intermittent hydrogen bon correlation function. (b) Continuous hydrogen bond correlation function.**

**Table 3. Timescales of hydrogen bond dynamics for three protonated forms of metformin hydrochloride.**

| Correlation function | Molecule | $a_1$ | $\tau_1$(ps) | $a_2$ | $\tau_2$(ps) | $<\tau>$(ps) |
|---|---|---|---|---|---|---|
| $C_{HB}(t)$ | MET1 | 0.49 | 0.51 | 0.51 | 10.81 | 5.8 |
| | MET2 | 0.41 | 0.33 | 0.59 | 8.40 | 5.1 |
| | MET3 | 0.39 | 0.38 | 0.61 | 7.14 | 4.5 |
| $S_{HB}(t)$ | MET1 | 0.85 | 0.01 | 0.15 | 0.16 | 0.032 |
| | MET2 | 0.86 | 0.01 | 0.14 | 0.23 | 0.041 |
| | MET3 | 0.88 | 0.01 | 0.12 | 0.42 | 0.059 |

## (E) Dynamical Features of Metformin in Aqueous Solution: Orientational Correlation

We investigate the rotational dynamics of MET in the form of first and second rank rotational time correlation functions as defined below,

$$C_1(t) = \langle P_1(\hat{\mu}_0 \cdot \hat{\mu}_t) \rangle; \text{ where } P_1(x) = x \tag{4}$$

$$C_2(t) = \langle P_2(\hat{\mu}_0 \cdot \hat{\mu}_t) \rangle; \text{ where } P_2(x) = \tfrac{1}{2}(3x^2 - 1) \tag{5}$$

Here, $P_1$ and $P_2$ are respectively the first and second rank Legendre polynomials and $\mu_t$ is the unit vector along N2—N4 axis (see **Figure 6**) of metformin. The rotational time constants are quite comparable for all three forms (**Figure 11**). These time constants are obtained by fitting the resultant curves with a multi exponential function and the average time constants are obtained by integrating the area under the curve with respect to time, as shown in **Table 4**.

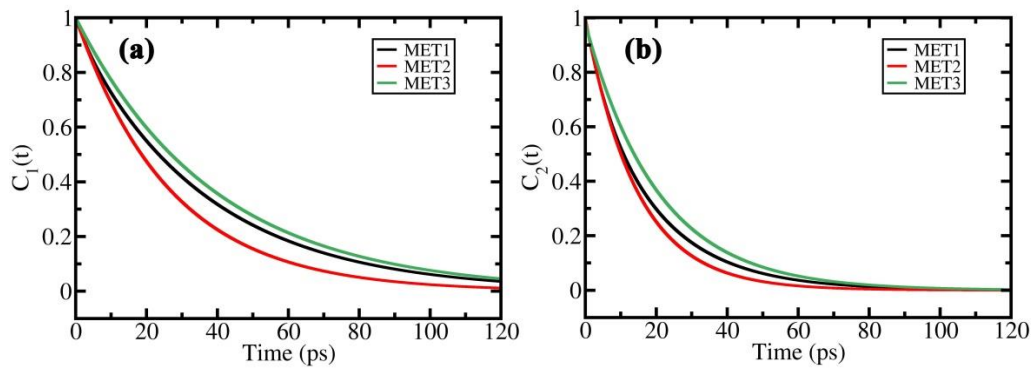

**Figure 11.** First (left, a) and second rank (right, b) rotational time correlation function for three different protonated forms of metformin. The three protonated forms show slightly different rotational dynamics with distinct time constants noted down in Table 4.

**Table 1.** Timescales of first and second rank orientational dynamics for three protonated forms of metformin hydrochloride.

| Correlation function | Molecule | $a_1$ | $\tau_1$(ps) | $a_2$ | $\tau_2$(ps) | $\langle\tau\rangle$(ps) |
|---|---|---|---|---|---|---|
| $C_1(t)$ | MET1 | 0.05 | 3.83 | 0.95 | 36.52 | 34.9 |
| | MET2 | --- | --- | 1.0 | 26.82 | 26.8 |
| | MET3 | --- | --- | 1.0 | 38.84 | 38.8 |
| $C_2(t)$ | MET1 | 0.18 | 5.99 | 0.82 | 19.3 | 16.9 |
| | MET2 | --- | --- | 1.0 | 14.5 | 14.5 |
| | MET3 | 0.03 | 0.21 | 0.97 | 20.4 | 19.8 |

Because of the asymmetric charge distribution of MET there exists a considerable contribution form dielectric friction in addition to Stokes' friction. This makes the dynamics non-Markovian and results in multiple timescales.[44] Hence we obtain a multi exponential fit. However, in some cases the time scales are fairly close to assume a single exponential behavior.

## (F) Solvation Dynamics of Metformin Hydrochloride

Solvation of small molecules is investigated in great details in several earlier studies.[45-47] It gives a measure how rapidly the surrounding dipolar solvent can respond to the changed charge distribution of the solute.[44, 48-49] Ideally it should be calculated from either non-equilibrium simulations or TDFSS experiments but that needs a proper identification of the excited states. However, under linear response approximation, time auto-correlation function of electrostatic interaction energy of metformin$^{(+)}$ with water molecules and ion equivalent to the non-equilibrium stokes shift response function measured by TDFSS experiments as described in Eq. (6).[44, 50-51] We calculate the time correlation function from a 5ns MD trajectory (at 2 fs trajectory dumping rate).

$$S(t) = \frac{v(t) - v(\infty)}{v(0) - v(\infty)} = \frac{\langle \delta E(0) \delta E(t) \rangle}{\langle \delta E(0)^2 \rangle} \quad (6)$$

Here, $\delta E(t)$ is the fluctuation in the interaction energy at time '$t$'. The resultant data is fitted to a tri-exponential function with an initial Gaussian component. The solvation of metformin hydrochloride is found to be ultrafast with an average time constant ranging from 130 to 160 fs with three distinctly different time constants (**Table 5**). One such exemplary plot is shown in **Figure 12** for MET3. Faster solvation dictates the increased stability of MET in liquid water.

**Table 2. Solvation timescales of three protonated forms of metformin hydrochloride in liquid water. Each of them shows ultrafast solvation with three different time scales. The fitting parameters are noted down in this table after fitting the resultant solvation energy time correlation function [Eq.(6)] to a tri-exponential function with an initial Gaussian decay [ $S(t) = a_g \exp(-t^2/\tau_g^2) + \sum_{i=1}^{2} a_i \exp(-t/\tau_i)$ ].**

| Molecule | $a_g$ | $\tau_g$ (fs) | $a_2$ | $\tau_2$ (fs) | $a_3$ | $\tau_3$ (fs) | $\langle\tau\rangle$ (fs) |
|---|---|---|---|---|---|---|---|
| MET1 | 0.36 | 10 | 0.48 | 115 | 0.16 | 598 | 154.5 |
| MET2 | 0.38 | 10 | 0.47 | 96 | 0.15 | 547 | 131.0 |
| MET3 | 0.35 | 10 | 0.44 | 100 | 0.21 | 530 | 158.5 |

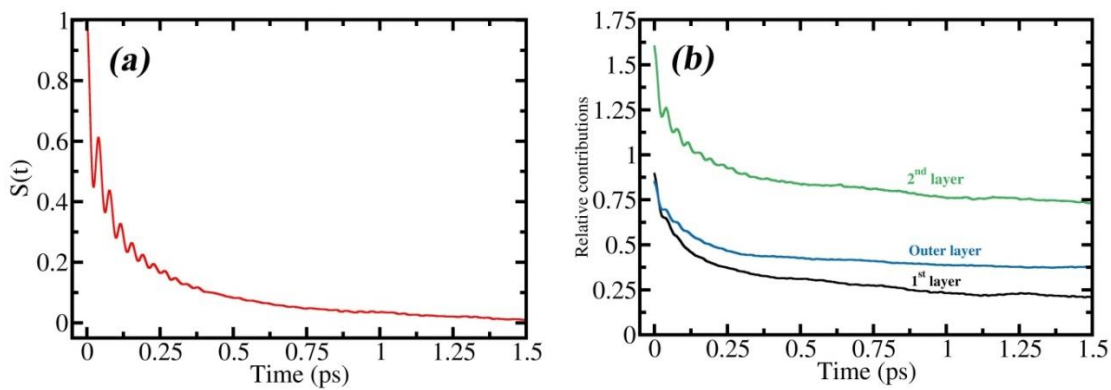

**Figure 12. (a) Total solvation energy relaxation for the tautomeric form of metformin hydrochloride (MET3) in water. (b) Layer wise decomposition of solvation energy relaxation that shows the role of second layer in the observed oscillations in total solvation energy relaxation.**

The solvation time correlation function is oscillatory in nature. We perform layer wise decomposition of the total solvation response to isolate the origin of these oscillations (**Figure 12**).[51] The Y-axis in **Figure 12(b)** is the relative contribution to the total solvation by each layer around metformin. There are negative-amplitude cross terms (not shown here) which neutralizes the large slow component arising from self-terms.

We observe that the *initial oscillations in the solvation energy time correlation function arise primarily due to the second layer* that has the highest amplitude as well. Because of the strong electrostatic interactions and long lived hydrogen bond formation ability of metformin$^{(+)}$ with the water molecules in the first shell, the ones in the second shell is more mobile and contributes to the oscillations that arises due to librational motion.

## (G) Interaction and Binding with DNA

### Theoretical Investigations

Because of the extended positive charge on the molecule, MET possesses a natural tendency to be attracted by DNA which is rich on negative charge on the phosphate backbone. In order to study the stability and pathway of formation of an outside groove bound state of MET3 we first prepare a docked structure with two model B-DNA dodecamer. The preferable docking position is observed to be the minor groove. We start our simulation by considering this as the initial configuration (**Supplementary Figure S4**).

In the case of metadynamics[52] simulation, 170 ns production run is found to be convergent. Two variables are used as order parameters to generate the free energy surface (see **Figure 13**). (i) The distance between the center of mass (COM) of MET3 and the COM of the four closest base pairs (C7, T8, A17 and D18) in the initial structure of GC rich DNA-MET complex. In case of the AT rich DNA, the distance between the COM of MET3 and COM of two closest base pair (D7 and D18) is chosen as the distance parameter. Let us call the closest base pairs the DNA reference groups in the initial structure; and (ii) the angle between a fixed vector, starting from the COMs of the reference groups and perpendicular to the DNA axis (that is in the body fixed frame description), and another vector joining the centre of mass of MET3 to the starting point of the previous vector. Hence this angle, if in between 0 to π/2 radian, indicates a minor groove bound state at the then a value between π/2 to π indicates major groove bound state at the initial configuration. However, these ranges shall be reversed when looked into the next turn of DNA. Hence, we need to look at the trajectories carefully to locate the structure of the complex inside a given free energy well.

The Gaussian hill height and width are taken to be 0.5 kJ/mol and 0.05 (in nm and radian units for distance and angle) respectively. A bias factor of 6.0 was assigned for well-tempered metadynamics[53]. We perform metadynamics simulations using PLUMED 2.0 package[54-55] patched with GROMACS 5.0.7[33].

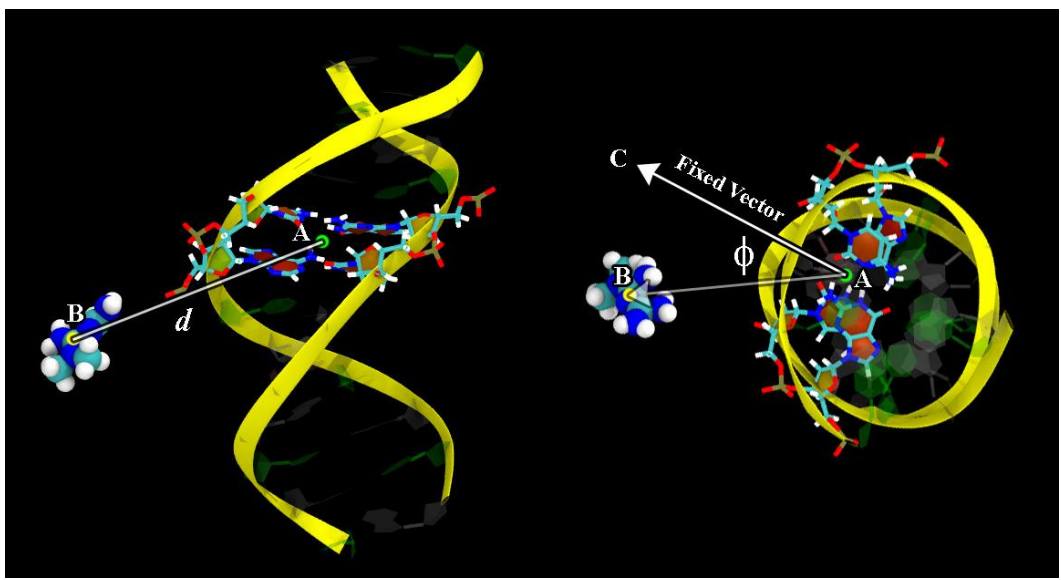

**Figure 13. Pictorial representation of the two chosen order parameters in order to generate the free energy surface of DNA-MET complex by the use of metadynamics simulation. The center of mass of four bases (in case of GC rich DNA, shown by green dot A) that are closest to the centre of mass of the drug (shown by yellow dot B) in the initial condition is chosen as a reference point from where the distance ($d$) and angle ($\phi$) are measured. In the above figure, AC vector is fixed with respect to the body fixed axis of DNA, but AB vector can change in both magnitude and direction as the drug explores different parts of the phase space during simulation.**

The free energy contours that are developed from the metadynamics simulation shows free energy minima regions (blue and indigo patches in **Figure 14**) in both AT and GC rich DNA-MET systems. In case of AT rich DNA there are several free energy wells in the free energy map (Figure 14 a). On careful examinations of the trajectories, we confirm that MET can bind to minor or major grooves and even with the phosphate backbone as well. On the other hand, there is lesser amount of stable regions in the free energy map of GC rich DNA. This suggests that MET can preferably bind to AT rich grooves/domains. The free energy cost to escape the deepest minima is obtained to be ~12-14 kJ/mol. However, this value is comparatively smaller than the binding free energies of other known groove binders (for example, Hoechst).[56-58]

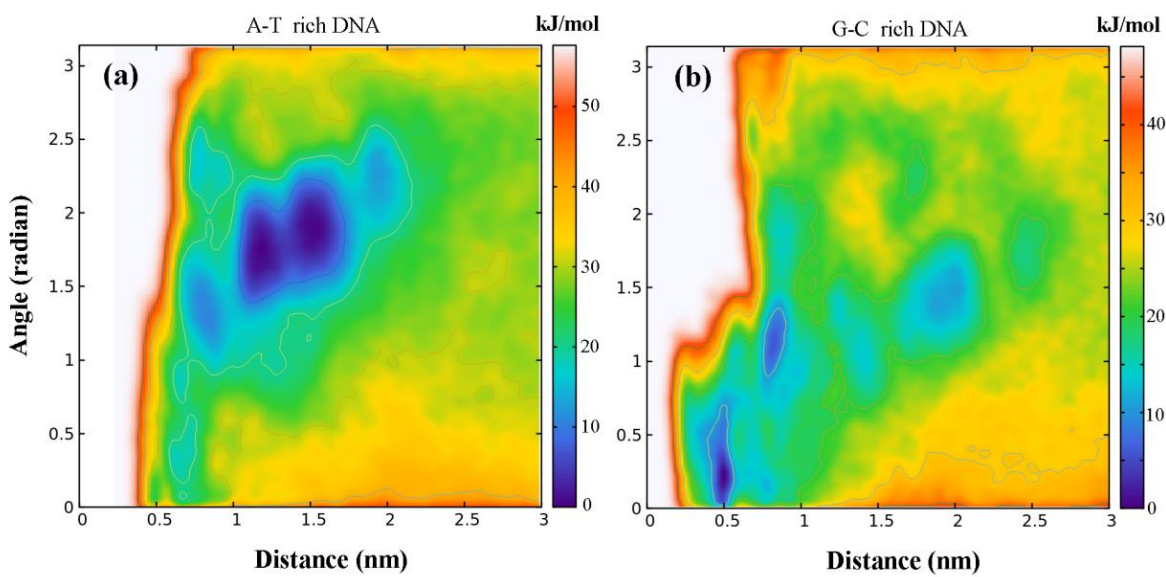

**Figure 14**. **Free energy contours of DNA-MET system with respect to two order parameters as described in Figure 13. The blue patches represent free energy minima. (a) In case if A-T rich DNA (A-T content ~80%) there are more blue patches which indicates that MET can be stabilized at multiple places on the surface of AT rich DNA. (b) In the case of G-C rich DNA (G-C content ~80%) the area consisting of blue patches is less. MET cannot bind at many position on the surface of the GC rich DNA. With both the model DNA it forms outer groove bound (major/minor) and backbone bound states as confirmed from the trajectory.**

Because of the low free energy of binding, MET can escape the free energy well (that is the minor groove bound state) in room temperature and explore other regions on the DNA surface. Few such instances are shown in **Figure 15** with the help of snapshots at different time points.

Let us start from an arbitrary time t=0 where MET first binds to the major groove of AT rich DNA. General considerations based on the relation between diffusion and entropy, such as Adam-Gibbs[59] or Rosenfeld scaling[60], indicate that the major groove bound state is entropically favourable because of greater translational and rotational degrees of freedom, although there are plenty of other factors like hydrogen bonding and arrangement of solvent molecules in the neighbourhood.[61] But the minor groove bound state is the global minimum. One interesting feature that is observed in our simulations is the restricted diffusion onto the DNA surface (**Figure 15**). It at first goes to a transient backbone bound state from the major groove. Then it slides into the minor groove pocket. This kind of feature appears to be a general characteristic of such small molecules in aqueous DNA systems.

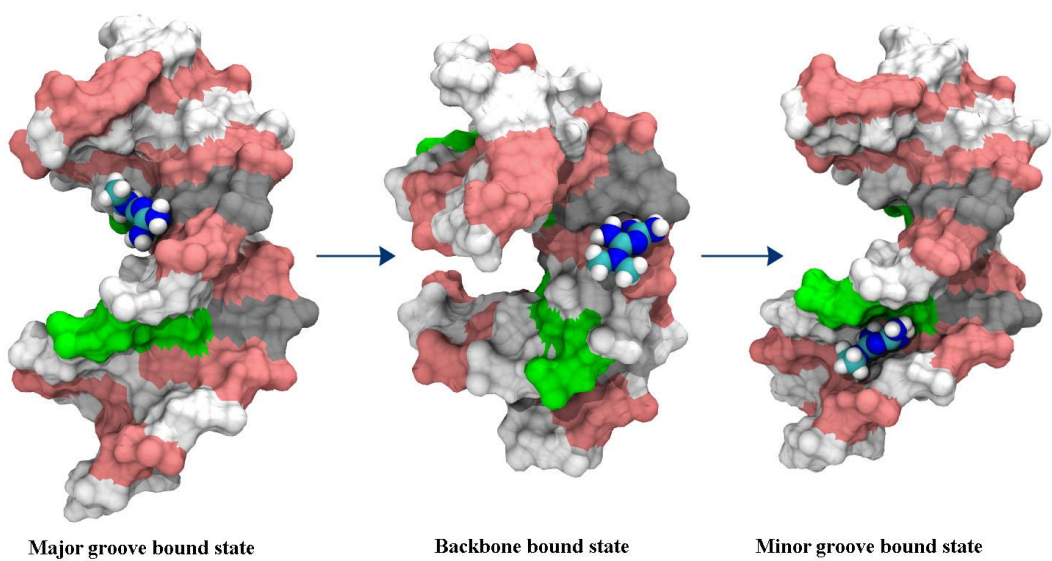

**Major groove bound state**  **Backbone bound state**  **Minor groove bound state**

**Figure 15. Diffusion of metformin on the surface of AT rich DNA. This sequence of figures depict how MET undergoes surface diffusion from major groove bound state to the more**

**stable minor groove bound state via a transient backbone bound state. The above figures are snapshots from metadynamics trajectories visualized using VMD[62].**

We set up another MD simulation (50 ns trajectory) without any bias potential with a B-DNA dodecamer and 10 MET3 molecules. This allows us to observe the inherent tendency of selective groove binding. We start with a random configuration where all MET3 molecules are randomly placed away from the DNA. In **Figure 16** we show the time evolution of the system. It is clear from the analysis of the trajectory that MET forms both major and minor groove bound state. But at the long time most of the *molecules tend to form a minor groove bound state like an extended chain that has significantly large residence time (~10-30 ns)*. The residence time of MET in the minor groove is high. The drug molecules form an outside bound state within few tens of ns which is also seen for other drug molecules like proflavin.[23] They become stable due to several strong hydrogen bond formations with the DNA base pairs that contribute to a higher Boltzman factor weightage. These results are consistent with the metadynamics results discussed above.

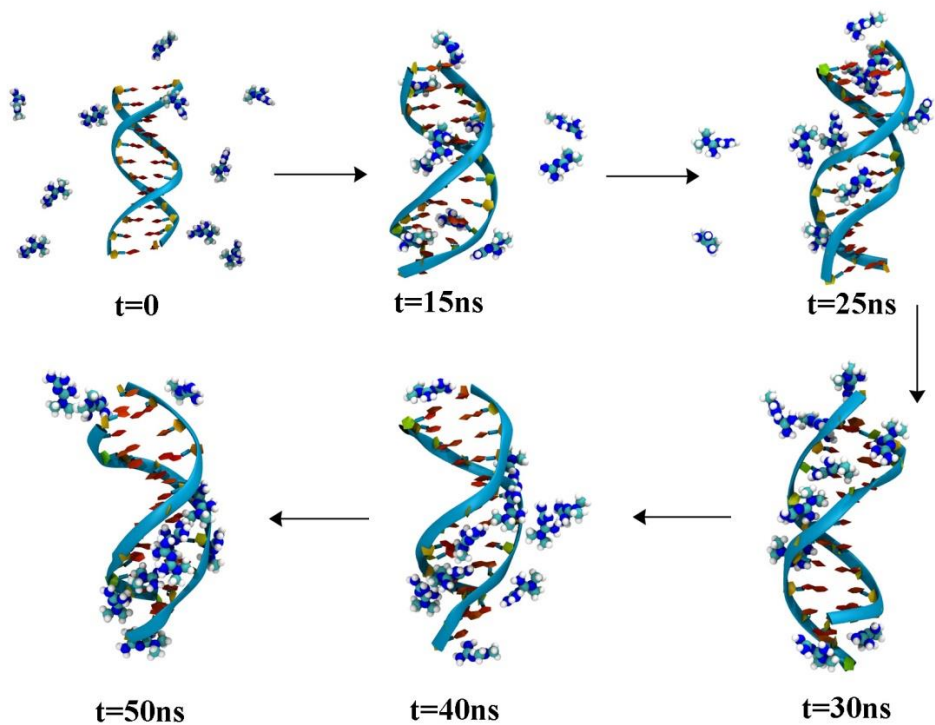

**Figure 16.** Time evolution of DNA-MET3 system where 10 MET and one GC-rich B-DNA is present. From the randomly distributed state the MET molecules form major/minor groove bound state. In the long time limit the minor groove bound state is found to be more favourable. The snaps are taken from a 50 ns long MD simulation using VMD.[62]

We carry out a quantitative analysis in order to observe the groove selectivity of individual drug molecules. We assign a value of '0' if it is bound to a minor groove and '1' if it goes to the major groove. Any states other than these two are kept blank. The results are shown in **Supplementary Figure S5**. The molecules switch the grooves but spend a longer time (~10-30 ns) in '0' state that is the minor groove. Some molecules fluctuate between the two states (**Figure S5** (e) and (g)) that indicates they are either at the two ends of the DNA or dangling from a backbone bond state.

In order to study interaction of MET with DNA we carry out several decisive experiments that are discussed in detail in the next section. The experimental studies aim to investigate the mode of interaction with DNA, that is, intercalating or outside groove binding drug.

Experimental Results on DNA Binding.

**Preparation of DNA solution.** We use commercially available Calf Thymus DNA (CT DNA) purchased from Sigma Aldrich Chemicals Pvt Ltd, India. Approximately 2mg of CT DNA is dissolved in 10 mM TrisHCl buffer in presence of 1mM EDTA. The EDTA is added to prevent in-situ denaturation of DNA strands. This stock solution is diluted further for spectroscopic studies. The UV-Vis absorption spectrum recorded for a 50 μg/mL dilution from this stock (**Supporting Figure S3**) shows a band at 260 nm characteristic of Nucleic Acids[63]. An $A_{260}/A_{280}$ ratio of 1.86 measured from this solution indicates absence of any protein impurities.

**Interaction of CT-DNA with Metformin – Electronic Absorption Spectra.** The absorption band due to nucleic acid is known to undergo changes when partial unfolding of the double helix structure of DNA occurs as a result of molecular intercalation into DNA. This occurs because of exposure of the nucleobases, which is a consequence of structural distortion due to intercalation. Electronic absorption band of small molecules also undergo changes when bound to DNA. Hyperchromicity or hypochromicity of the absorption band can indicate the mode of interaction with DNA.[64]

We record the electronic absorption spectrum of a 100 μM solution of Metformin and study the effect of addition of DNA to this solution. The electronic absorption spectrum of 100 μM in presence of increasing concentration of DNA (25 μg/mL to 200 μg/mL) is shown in **Figure 17**. We note that no change in the position of the absorption band of pure metformin, but a gradual hyperchromicity is observed in the presence of DNA. This type of behaviour is documented in literature and suggests non – intercalative mode of interaction.[18, 29, 65]

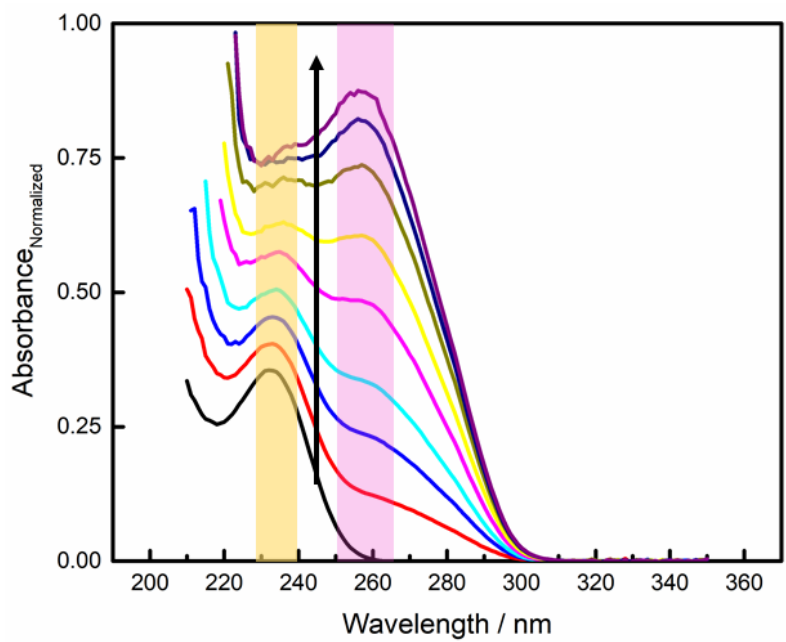

**Figure 17. UV – visible absorption of Metformin (100 µM) in solution (black trace) and in the presence of increasing concentration of CT-DNA (25 µg/mL – red trace to 200 µg/mL – purple trace). The absorption band of pure metformin is highlighted in orange while that of pure CT DNA is highlighted in pink. The increase in the absorption (hyperchromism) of metformin in the presence of DNA (as shown in the arrow) suggests non – intercalation mode of interaction.**

**Structural change of DNA on interaction with Metformin: Circular Dichroism study.** Circular Dichroism spectroscopy is useful for studying changes in the structure of double – helix DNA strands on interaction with small molecules[66]. The circular dichroism spectrum of double helix DNA is characterized by a positive and negative bands centred around 260 nm and a weak positive band around 220 nm.[67] Changes in helicity because of intercalation results in alteration of the circular dichroism spectra of the DNA strands[68].

We record the circular dichroism data in a JASCO J815 instrument. A 1 mm path length quartz cuvette (Hellma) is used for the measurements. We measure the CD spectra of 100 μg/mL DNA in buffer as well as in the presence of increasing concentration of metformin (25 μg/mL to 300 μg/mL) as shown in **Figure 18**. No particular change in the circular dichroism spectra is seen in the DNA samples incubated with metformin in the dichroism cantered around 260 nm. This rules out any significant structural distortion of the double helix which in turn excludes intercalation as a possible mode of DNA – metformin interaction. Thus the circular dichroism data supports the indication obtained from the UV – vis absorption spectra that metformin does not intercalate into the DNA helix. However, we also notice an increasing distortion in the positive band at 220 nm. As metformin also absorbs ~230 nm, we speculate this to be the induced circular dichroism (ICD) of metformin in the presence of DNA. Small molecules like metformin, on binding to helical molecular moiety of DNA can exhibit ICD.[69-70]

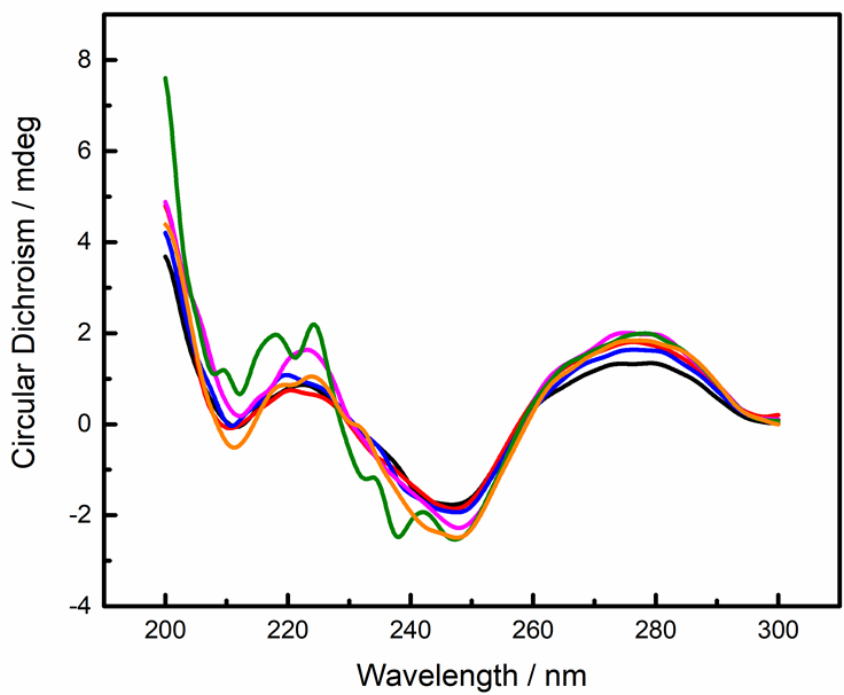

**Figure 18. Circular dichroism spectra of CT DNA (black trace) and DNA incubated with increasing stoichiometric ratios of Metformin (DNA:Met = 1:0.25 (red), 1:0.5 (blue), 1:1 (orange), 1:2 (pink), 1:3 (green)). The characteristic negative and positive bands of the Cotton effect couplet centred on 275 nm do not undergo significant change in the presence of metformin as seen in the spectra. However the peak at 220 nm splits at a higher drug concentration. We speculate this as the induced circular dichroism (ICD) of metformin (natural UV absorption ~230 nm) by DNA.**

**Photoluminescence studies on DNA – metformin interaction.** Metformin in itself is a non – luminescent molecule that makes it unsuitable for photoluminescence studies. The data obtained from UV – visible and circular dichroism spectra indicate the presence of DNA – metformin interaction but rule out intercalation. This leads us to the conclusion that the possible mode of interaction may be groove binding. Acridine Orange and Hoechst are two standard fluorophores that shown enhancement in fluorescence on binding with DNA. While Hoechst is known to bind to DNA grooves, Acridine Orange intercalates within the DNA base pairs.

We study the fluorescence of DNA in the presence of Acridine Orange and Hoechst 34580 stain solution and note the changes in fluorescence when these are incubated with increasing concentrations of metformin. Fluorescence spectra are recorded in a 10 mm path length quartz cuvette (Hellma) on a Horiba Fluoromax fluorimeter. The Hoechst samples are excited at 390 nm and emission is collected from 400 nm to 650 nm. A working concentration of 50 μg/mL 0.02 μM Hoechst dye is used. The Acridine Orange samples are excited at 460 nm and emission data is collected over 470 nm to 700 nm. A working concentration of 5 μM Acridine Orange is used for the experiments. DNA concentrations of both fluorescence experiments are 50 μg/mL.

We see that for the DNA – Hoechst complex, there is a gradual decrease of fluorescence on incubation with metformin (**Figure 19 (A)**). As it is known that the Hoechst molecule shows intense fluorescence only when bound to the DNA, we infer from this observation that metformin in capable of partially replacing the bound Hoechst molecules in the minor grooves of the DNA. However, we also note that the decrease in fluorescence in only seen in samples that have been incubated with metformin for over a week. This may indicate kinetic sluggishness of the competitive interaction between Metformin and Hoechst molecules for binding to the minor groove of DNA double helix.

There is no significant change in fluorescence of the DNA – Acridine Orange complex upon incubation with metformin (**Figure 19(B)**), apart from a slight increase in intensity of fluorescence which may be due to increase in concentration of the DNA – AO bound complex with time.

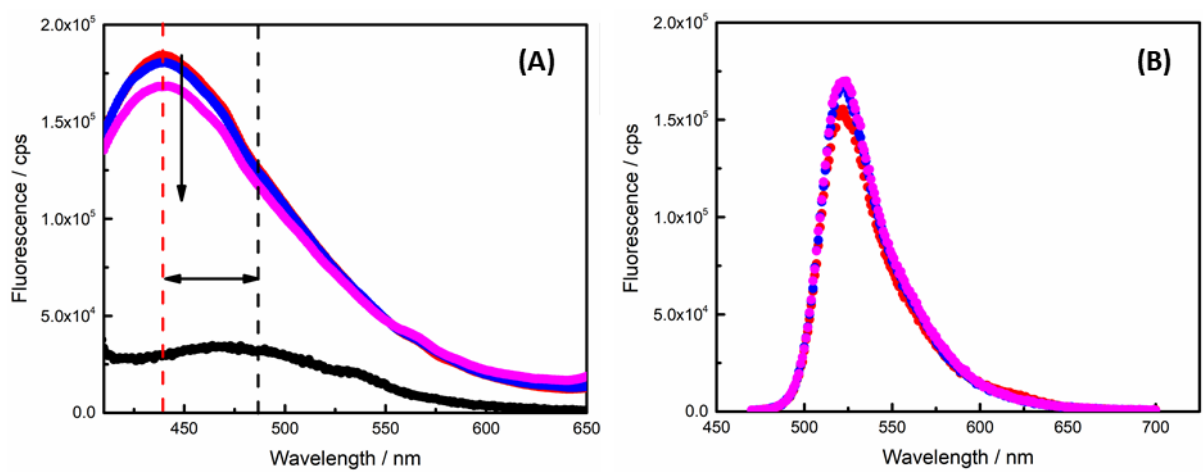

**Figure 19. (A) Fluorescence spectra of Hoechst (black trace), DNA – Hoechst (red trace), DNA – Hoechst – Metformin (blue trace – 10 μM, pink trace – 20 μM). The characteristic emission peak of DNA bound Hoechst is marked in red dashed line, the emission of unbound Hoechst is marked in black dashed line. The shift in the fluorescence peak on Hoechst binding to DNA is indicated by the double - sided arrow. The decrease in fluorescence with increasing concentration of DNA is shown by the arrow pointing downwards. (B) Fluorescence spectra of DNA – Acridine Orange (red trace) and DNA – Acridine Orange – Metformin (blue trace – 10 μM, pink trace 20 μM). No decrease of**

**fluorescence in the presence of metformin establishes the inability of metformin in replacing the interacalating Acridine Orange molecules in the double stranded DNA helix.**

- ## SUMMARY AND CONCLUSIONS

Here we have carried out experimental and theoretical investigations to explore and to understand characteristic properties of metformin hydrochloride. To the best of our knowledge, such studies have not been directed to MET earlier. This prototype study can well be followed in cases of other biologically relevant small molecules. We summarize our results with the help of the following points.

i. We find several interesting characteristics of this important molecule not discussed in the literature. First, the protonated nitrogen rich molecule exhibits multiple protonated forms. It exhibits both a planar and non-planar form with the non-planar one being slightly more stable. The delocalised positive charge helps it to interact with charged surfaces of biomolecules.

ii. We establish the abundance of the tautomer in a sample of metformin hydrochloride with the help of UV-Vis and IR spectroscopic studies. As adaptability could be a key feature of biological activity, this might hold the key to the diverse activity of MET.

iii. We have generated a sustainable AMBER based force field for metformin for classical MD simulation. We have validated the force field and chosen the most abundant protonated form with respect to ab-initio calculations, diffusion and melting studies. The results from experiments and simulations are in good agreement.

iv. The barrier of proton transfer between two mono protonated forms of metformin is calculated to be ~4-6 kcal/mol. This allows us to assume a possible co-existence of the two forms in aqueous solution and run the simulations accordingly.

v. We have observed solvation dynamics, hydrogen bond lifetime and rotational dynamics of different protonated forms of metformin. The solvation is ultrafast with an average time constant of 130-160 fs. The hydrogen bond lifetimes has direct consequence on the diffusion coefficient of different protonated forms.

vi. Because of the reported cancer preventing and tumour supressing roles of metformin, its interaction with DNA draws avid attention. We perform several experimental and metadynamics simulation studies to understand the mode and pathway of binding. We observe that MET can bind to grooves and backbone of DNA within a few tens of ns MD timescale. *The minor groove bound state is the most stable one*. The extent of binding varies with the base pair ratios. We find that MET prefers AT rich domains over GC rich ones.

vii. Our experiments also suggest a minor groove bound state as it can slowly replace a known strong minor groove binder Hoechst. This groove binding property can be exploited to encumber the normal DNA replication scheme that would finally result in a reduced cell multiplication. *However, we do not observe any signature of intercalation.* The CD spectra at higher drug concentrations show splitting and increase of the 220nm band that allows us to speculate the CD peak of the drug induced by DNA as the drug molecules gradually assemble to the grooves (**Figure 16**).

One of our primary objectives has been to search for the characteristic features of small drug molecules (for example steroids, neurotransmitters, vitamins etc.) that allowed them to perform the function they are used for. Although one hopes to find certain universal chemical characteristics, we are aware that such a goal is difficult and would require the study of a large number of molecules. Many small molecules serve as powerful drugs because of their strong effects on the human body. Often the microscopic aspects of such effects remain unclear. Metformin is not only a small molecule but characterized by unique features like five N atoms and in its hydrochloride form electron deficiency is delocalized over these five N atoms. In this particular case, the appearance of tautomeric forms on protonation seems to offer a clue – the molecule is adaptive to different environments. The protonation also makes it largely planar that might help it to slide through the cell membranes. With water molecules and ions all around, acquired polarity certainly could help, for example, in binding to DNA. Many of the above are not unexpected. But which of the above properties aids in its role as an anti-diabetic drug is unanswered. A quantitative study and molecular level understanding of the drug action and mechanism of metformin hydrochloride still remain elusive.

**ACKNOWLEDGMENT**

We thank Prof P. Balaram for insightful discussions and guidance. R. N. Samajdar thanks Mr. Rupak Saha and Mr. Sunit Pal for help with the experiments, and Mr. Arka Baksi for providing the Hoechst sample for analysis. We also thank the Department of Science and Technology (DST, India) for partial support. B. Bagchi thanks Sir J C Bose fellowship. S. Mondal thanks UGC, India for providing research fellowship and S. Mukherjee thanks DST, India for providing INSPIRE fellowship.

# Supporting Information

*In the main text we report and discuss various aspects of structure, dynamics and DNA binding energetics of anti-diabetic drug metformin hydrochloride. Here, under supplementary figures, we provide the NMR and UV spectrum of metformin and UV spectrum of DNA which may be regarded as the test for purity. The initial configurations taken for metadynamics and residence time plots of metformin in DNA grooves are also provided. System size dependence of diffusion constant is discussed. We also provide the GAFF based and somewhat refined force field parameters of various mon-protonated forms of metformin in order to ensure the reproducibility of our MD and metadynamics simulations.*

**SUPPORTING FIGURES:**

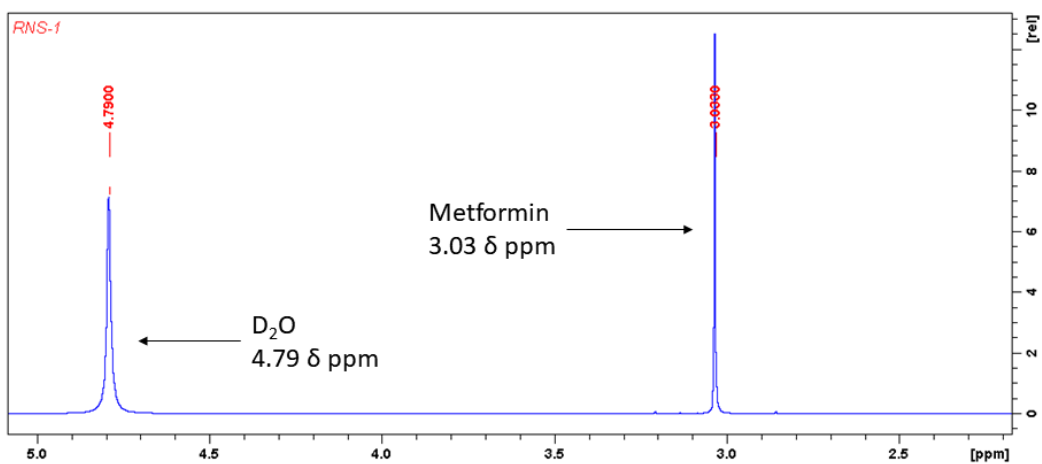

**Supporting Figure S1:** $^1$H NMR spectrum of metformin crystals dissolved in D$_2$O, showing the characteristic methyl protons at 3.03 δ ppm. The solvent peak is seen at 4.79 δ ppm.

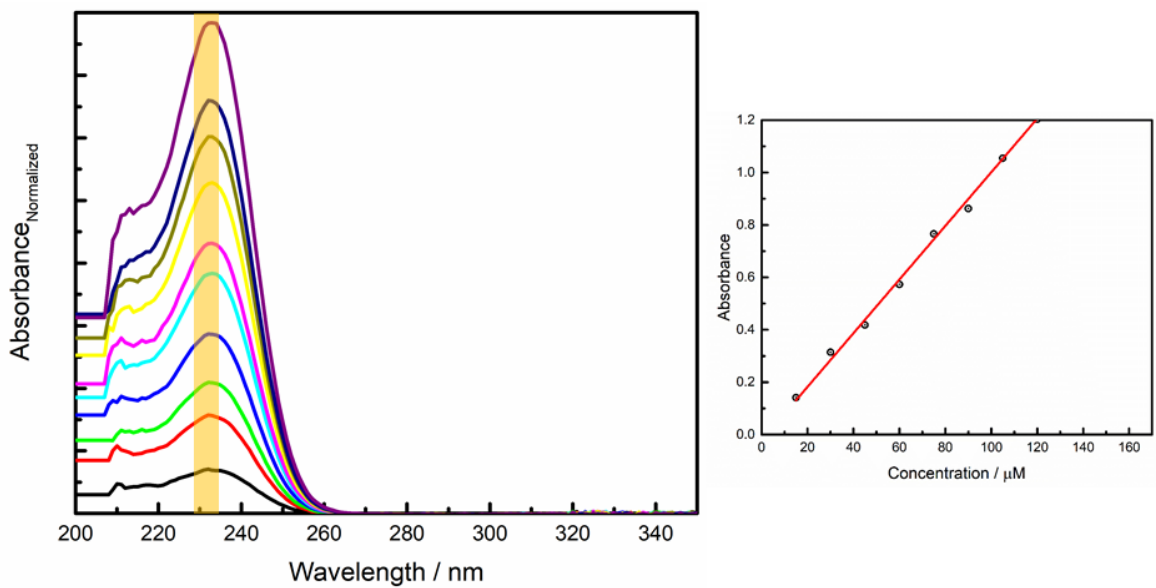

**Supporting Figure S2:** (Left) UV – visible absorption spectrum of Metformin showing characteristic absorption band centred around 230 nm (highlighted in orange). No noticeable shift of this band occurs in the concentration range shown (15 μM : black trace to 150 μm : purple trace, in increments of 15 μM every trace). (Right) Absorption – concentration plot of Metformin showing conformity with Lambert Beer approximation in the concentration range used with ε calculated to be 10240 (±230) M$^{-1}$cm$^{-1}$.

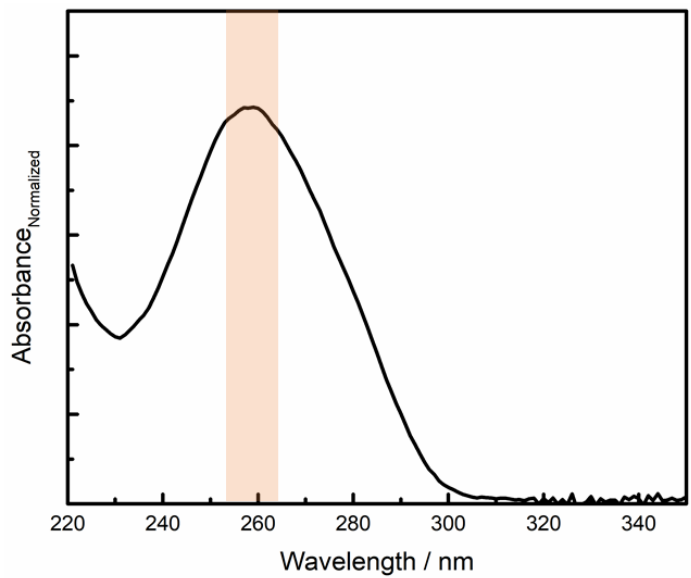

**Supporting Figure S3**: Electronic absorption spectrum of calf thymus DNA with the nucleic acid absorption band at 260 nm highlighted in light red.

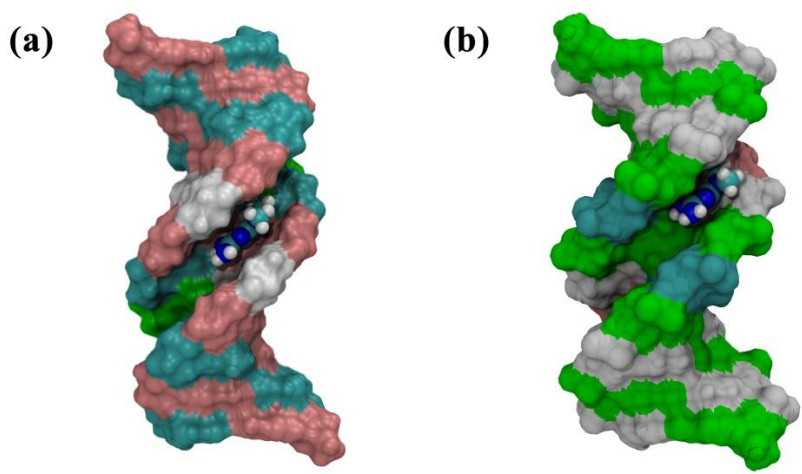

**Supporting Figure S4: The initial structures of metadynamics simulation prepared using Schrodinger software package. The complex assumes a minor groove bound state with both (a) AT rich and (b) GC rich DNA.**

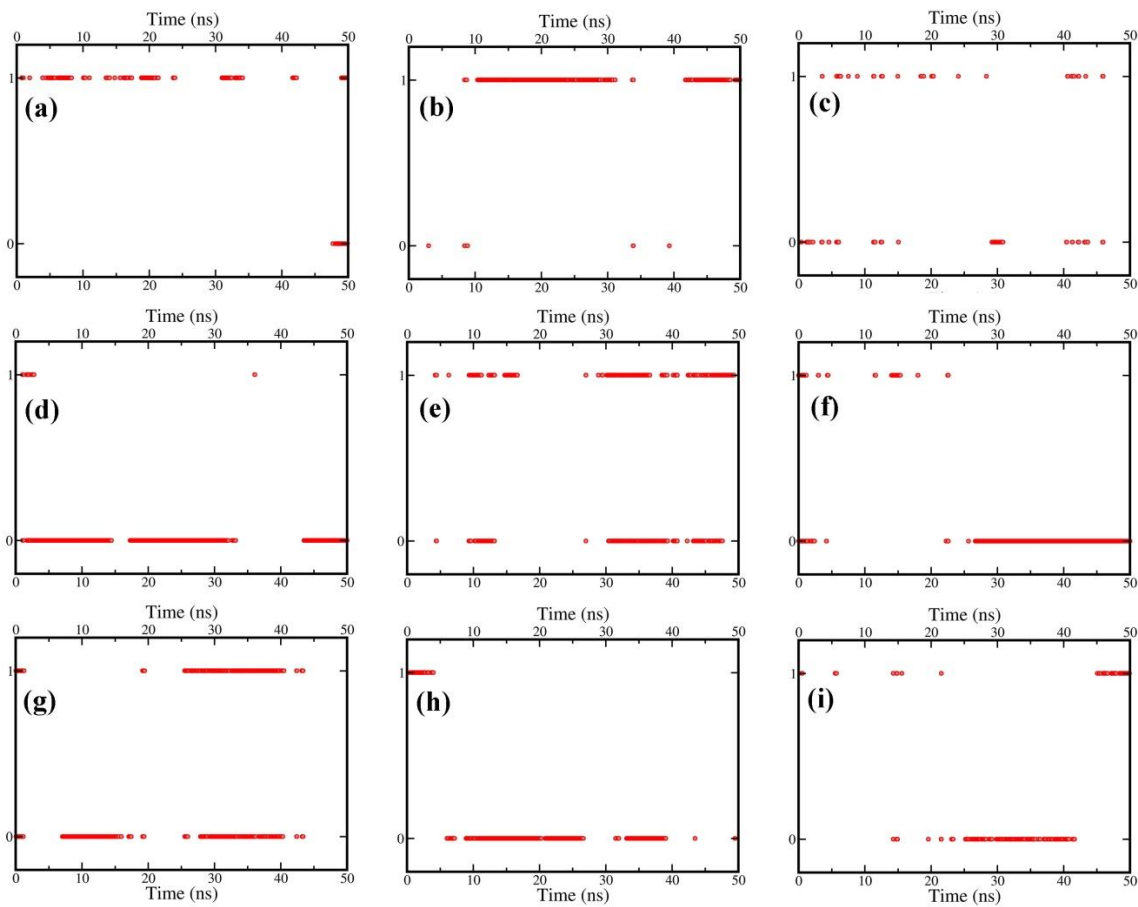

**Supporting Figure S5.** Groove selectivity of metformin hydrochloride is represented in this figure. The state '0' represents a minor groove bound state and '1' represents a major groove bound state. The blank regions are any state other that these two. It is clear that the drug molecule has long residence time when bound to a minor/major groove. Statistically the minor groove is more populated at a given time. MET can also switch grooves that is described by the jumps between '0' and '1' state.

# SYSTEM SIZE DEPENDENCE OF DIFFUSION

**Table S1:** We calculate diffusion constant (in cm$^2$/s) for centre of mass diffusion of metformin molecules. In the main text, only the values for the largest system are provided. Here we provide the rest.

| System | MET1 | MET2 | MET3 |
|---|---|---|---|
| 500 water + 1MET | $0.71(\pm 0.11) \times 10^{-5}$ | $1.67(\pm 0.60) \times 10^{-5}$ | $1.81(\pm 0.54) \times 10^{-5}$ |
| 1440 water + 3 MET | $1.61(\pm 0.24) \times 10^{-5}$ | $1.18(\pm 0.09) \times 10^{-5}$ | $2.0(\pm 0.3) \times 10^{-5}$ |
| 2880 water + 6 MET | $1.90(\pm 0.40) \times 10^{-5}$ | $0.89(\pm 0.17) \times 10^{-5}$ | $1.07(\pm 0.26) \times 10^{-5}$ |
| 4320 water + 9 MET | $1.13(\pm 0.35) \times 10^{-5}$ | $1.25(\pm 0.45) \times 10^{-5}$ | $1.47(\pm 0.53) \times 10^{-5}$ |

# TOPOLOGY AND FORCE-FIELD OF METFORMIN HYDROCHLORIDE

We have generated sustainable force fields for MD simulation of three different protonated forms of metformin hydrochloride using GAFF based atom type description and relevant quantum chemical calculations. Below we provide the force field details (partial charges, force constants etc.) and atomic coordinates of these three forms in order to ensure the reproducibility of our results and future uses/refinements. We prepare the files in GROMACS compatible format (that is *.gro* for coordinates and *.itp* for topology) so that these can be directly used. For the detailed meaning of each column of data represented below, please refer to any standard GROMACS manual[71].

**Coordinates of MET1 (.gro format)**

```
resid resname  atomtype  resnr      x        y        z

    1MET1       N2       1    -0.284    0.107   -0.018
    1MET1       H1       2    -0.257    0.203   -0.033
    1MET1       H3       3    -0.382    0.086   -0.004
    1MET1       C4       4    -0.191    0.011   -0.004
    1MET1       N5       5    -0.223   -0.112    0.035
    1MET1       H6       6    -0.147   -0.181    0.035
    1MET1       H7       7    -0.318   -0.134    0.063
    1MET1       N8       8    -0.062    0.043   -0.030
    1MET1       H9       9    -0.046    0.133   -0.075
    1MET1      C10      10     0.052   -0.040   -0.012
    1MET1      N11      11     0.034   -0.168    0.002
    1MET1      H12      12     0.122   -0.215    0.024
    1MET1      N13      13     0.170    0.030   -0.021
    1MET1      C14      14     0.187    0.158    0.050
    1MET1      H15      15     0.092    0.214    0.054
    1MET1      H16      16     0.221    0.140    0.154
    1MET1      H17      17     0.261    0.218   -0.003
    1MET1      C18      18     0.292   -0.052   -0.021
    1MET1      H19      19     0.313   -0.095    0.079
    1MET1      H20      20     0.284   -0.132   -0.096
    1MET1      H21      21     0.376    0.014   -0.048
```

**Topology and Force-Field parameters of MET1 (.itp format)**

```
[ moleculetype ]
; Name            nrexcl
MET1              3

[ atoms ]
; nr     type   resnr   residue  atom  cgnr   charge      mass

   1      nh      1     MET1     N2     1   -0.900266   14.0100   ; qtot -0.9003
   2      hn      1     MET1     H1     2    0.461612    1.0080   ; qtot -0.4387
   3      hn      1     MET1     H3     3    0.461612    1.0080   ; qtot  0.0230
   4      cz      1     MET1     C4     4    0.925937   12.0100   ; qtot  0.9489
   5      nh      1     MET1     N5     5   -0.900266   14.0100   ; qtot  0.0486
   6      hn      1     MET1     H6     6    0.461612    1.0080   ; qtot  0.5102
   7      hn      1     MET1     H7     7    0.461612    1.0080   ; qtot  0.9719
   8      ne      1     MET1     N8     8   -0.666727   14.0100   ; qtot  0.3051
   9      hn      1     MET1     H9     9    0.387913    1.0080   ; qtot  0.6930
  10      c2      1     MET1    C10    10    0.716845   12.0100   ; qtot  1.4099
  11      n2      1     MET1    N11    11   -0.859902   14.0100   ; qtot  0.5500
  12      hn      1     MET1    H12    12    0.383606    1.0080   ; qtot  0.9336
  13      nh      1     MET1    N13    13   -0.174984   14.0100   ; qtot  0.7586
  14      c3      1     MET1    C14    14   -0.315343   12.0100   ; qtot  0.4433
  15      h1      1     MET1    H15    15    0.145347    1.0080   ; qtot  0.5886
  16      h1      1     MET1    H16    16    0.145347    1.0080   ; qtot  0.7340
  17      h1      1     MET1    H17    17    0.145347    1.0080   ; qtot  0.8793
```

```
    18          c3      1    MET1    C18    18  -0.315343   12.0100  ; qtot 0.5640
    19          h1      1    MET1    H19    19   0.145347    1.0080  ; qtot 0.7093
    20          h1      1    MET1    H20    20   0.145347    1.0080  ; qtot 0.8547
    21          h1      1    MET1    H21    21   0.145347    1.0080  ; qtot 1.0000

 [ bonds ]
 ;   ai     aj  funct     c0         c1
      1      2    1    0.10121 338569.28
      1      3    1    0.10121 338569.28
      1      4    1    0.13390 408358.40
      4      5    1    0.13390 408358.40
      4      8    1    0.12920 480406.88
      5      6    1    0.10121 338569.28
      5      7    1    0.10121 338569.28
      8      9    1    0.10230 322670.08
      8     10    1    0.12915 480406.88
     10     11    1    0.12817 497142.88
     10     13    1    0.13872 348276.16
     11     12    1    0.10230 322670.08
     13     14    1    0.14640 273298.88
     13     18    1    0.14640 273298.88
     14     15    1    0.10969 276646.08
     14     16    1    0.10969 276646.08
     14     17    1    0.10969 276646.08
     18     19    1    0.10969 276646.08
     18     20    1    0.10969 276646.08
     18     21    1    0.10969 276646.08

 [ pairs ]
 ;   ai     aj funct
      1      6    1
      1      7    1
      1      9    1
      1     10    1
      2      5    1
      2      8    1
      3      5    1
      3      8    1
      4     11    1
      4     13    1
      5      9    1
      5     10    1
      6      8    1
      7      8    1
      8     12    1
      8     14    1
      8     18    1
      9     11    1
```

```
         9         13         1
        10         15         1
        10         16         1
        10         17         1
        10         19         1
        10         20         1
        10         21         1
        11         14         1
        11         18         1
        12         13         1
        14         19         1
        14         20         1
        14         21         1
        15         18         1
        16         18         1
        17         18         1

[ angles ]
;    ai        aj        ak     funct       c0           c1
      1         4         5         1    120.14005  610.780320
      1         4         8         1    123.46005  600.822400
      2         1         3         1    115.12005  335.305760
      2         1         4         1    121.15005  408.525760
      3         1         4         1    121.15005  408.525760
      4         5         6         1    121.15005  408.525760
      4         5         7         1    121.15005  408.525760
      4         8         9         1    110.80005  441.746720
      4         8        10         1    118.18005  595.048480
      5         4         8         1    123.46005  600.822400
      6         5         7         1    115.12005  335.305760
      8        10        11         1    113.82005  655.549120
      8        10        13         1    123.46005  600.822400
      9         8        10         1    110.80005  441.746720
     10        11        12         1    110.80005  441.746720
     10        13        14         1    123.71005  522.163200
     10        13        18         1    123.71005  522.163200
     11        10        13         1    124.27005  600.738720
     13        14        15         1    109.79005  414.801760
     13        14        16         1    109.79005  414.801760
     13        14        17         1    109.79005  414.801760
     13        18        19         1    109.79005  414.801760
     13        18        20         1    109.79005  414.801760
     13        18        21         1    109.79005  414.801760
     14        13        18         1    114.51005  529.276000
     15        14        16         1    108.46005  328.360320
     15        14        17         1    108.46005  328.360320
     16        14        17         1    108.46005  328.360320
```

```
    19      18      20     1    108.46005 328.360320
    19      18      21     1    108.46005 328.360320
    20      18      21     1    108.46005 328.360320

[ dihedrals ]
;   ai      aj      ak     al   funct    c0       c1      c2

     1       4       5      6     1    180.00   17.3636   2
     1       4       5      7     1    180.00   17.3636   2
     1       4       8      9     1    180.00   17.3636   2
     1       4       8     10     1    180.00   17.3636   2
     2       1       4      5     1    180.00   17.3636   2
     2       1       4      8     1    180.00   17.3636   2
     3       1       4      5     1    180.00   17.3636   2
     3       1       4      8     1    180.00   17.3636   2
     4       8      10     11     1    180.00   17.3636   2
     4       8      10     13     1    180.00   17.3636   2
     5       4       8      9     1    180.00   17.3636   2
     5       4       8     10     1    180.00   17.3636   2
     6       5       4      8     1    180.00   17.3636   2
     7       5       4      8     1    180.00   17.3636   2
     8      10      11     12     1    180.00   17.3636   2
     8      10      13     14     1    180.00   17.3636   2
     8      10      13     18     1    180.00   17.3636   2
     9       8      10     11     1    180.00   17.3636   2
     9       8      10     13     1    180.00   17.3636   2
    10      13      14     15     1    180.00   1.00000   6
    10      13      18     19     1    180.00   1.00000   6
    11      10      13     14     1    180.00   17.3636   2
    11      10      13     18     1    180.00   17.3636   2
    12      11      10     13     1    180.00   17.3636   2
    14      13      18     19     1    180.00   1.00000   6
    15      14      13     18     1    180.00   1.00000   6
; improper dihedrals
     4       5       1      8     4      0.00   4.60240   2
    10      13       8     11     4      0.00   4.60240   2
    13      10      14     18     4      0.00   4.60240   2
     1       4       2      3     4      0.00   4.60240   2
------------------------------------------------------------------------
```

**Coordinates of MET2 (.gro format)**

resid resname  atomtype  resnr     x       y       z

    1MET2    N2      1    -0.282   0.112  -0.005
    1MET2    H1      2    -0.252   0.196   0.044

```
    1MET2      H3       3   -0.380    0.089    0.012
    1MET2      C4       4   -0.196    0.004   -0.002
    1MET2      N5       5   -0.226   -0.122   -0.000
    1MET2      H6       6   -0.326   -0.136    0.012
    1MET2      N7       7   -0.061    0.046   -0.011
    1MET2      H8       8   -0.047    0.142   -0.039
    1MET2      C9       9    0.050   -0.034   -0.001
    1MET2     N15      10    0.035   -0.167    0.006
    1MET2     H16      11    0.113   -0.226    0.032
    1MET2     H17      12   -0.062   -0.200    0.006
    1MET2     N10      13    0.170    0.025   -0.001
    1MET2     C11      14    0.184    0.171    0.009
    1MET2     H12      15    0.115    0.210    0.085
    1MET2     H13      16    0.286    0.193    0.040
    1MET2     H14      17    0.165    0.219   -0.088
    1MET2     C18      18    0.291   -0.059   -0.007
    1MET2     H19      19    0.313   -0.103    0.092
    1MET2     H20      20    0.278   -0.138   -0.081
    1MET2     H21      21    0.374    0.005   -0.037
```

**Topology and Force-Field parameters of MET2 (.itp format)**

```
[ moleculetype ]
; Name                 nrexcl
MET2              3

[ atoms ]
;   nr      type   resnr residue  atom   cgnr    charge        mass

    1       nh     1     MET2    N2      1   -0.876652   14.0100   ; qtot -0.8767
    2       hn     1     MET2    H1      2    0.410536    1.0080   ; qtot -0.4661
    3       hn     1     MET2    H3      3    0.410536    1.0080   ; qtot -0.0556
    4       cz     1     MET2    C4      4    0.734160   12.0100   ; qtot  0.6786
    5       n2     1     MET2    N5      5   -0.829116   14.0100   ; qtot -0.1505
    6       hn     1     MET2    H6      6    0.389987    1.0080   ; qtot  0.2395
    7       ne     1     MET2    N7      7   -0.493060   14.0100   ; qtot -0.2536
    8       hn     1     MET2    H8      8    0.373652    1.0080   ; qtot  0.1200
    9       c2     1     MET2    C9      9    0.509524   12.0100   ; qtot  0.6296
   10       nh     1     MET2    N15    10   -0.699288   14.0100   ; qtot -0.0697
   11       hn     1     MET2    H16    11    0.397876    1.0080   ; qtot  0.3282
   12       hn     1     MET2    H17    12    0.397876    1.0080   ; qtot  0.7260
   13       nh     1     MET2    N10    13   -0.024315   14.0100   ; qtot  0.7017
   14       c3     1     MET2    C11    14   -0.314821   12.0100   ; qtot  0.3869
   15       h1     1     MET2    H12    15    0.154654    1.0080   ; qtot  0.5415
   16       h1     1     MET2    H13    16    0.154654    1.0080   ; qtot  0.6962
   17       h1     1     MET2    H14    17    0.154654    1.0080   ; qtot  0.8509
   18       c3     1     MET2    C18    18   -0.314821   12.0100   ; qtot  0.5360
   19       h1     1     MET2    H19    19    0.154654    1.0080   ; qtot  0.6907
   20       h1     1     MET2    H20    20    0.154654    1.0080   ; qtot  0.8453
   21       h1     1     MET2    H21    21    0.154654    1.0080   ; qtot  1.0000

[ bonds ]
;   ai     aj funct           c0           c1           c2           c3
```

```
     1         2      1     0.10121 338569.280000
     1         3      1     0.10121 338569.280000
     1         4      1     0.13390 408358.400000
     4         5      1     0.12820 497142.880000
     4         7      1     0.12920 480406.880000
     5         6      1     0.10230 322670.080000
     7         8      1     0.10230 322670.080000
     7         9      1     0.12915 480406.880000
     9        10      1     0.13872 348276.160000
     9        13      1     0.13872 348276.160000
    10        11      1     0.10121 338569.280000
    10        12      1     0.10121 338569.280000
    13        14      1     0.14640 273298.880000
    13        18      1     0.14640 273298.880000
    14        15      1     0.10969 276646.080000
    14        16      1     0.10969 276646.080000
    14        17      1     0.10969 276646.080000
    18        19      1     0.10969 276646.080000
    18        20      1     0.10969 276646.080000
    18        21      1     0.10969 276646.080000

[ pairs ]
;    ai       aj funct
     1         6      1
     1         8      1
     1         9      1
     2         5      1
     2         7      1
     3         5      1
     3         7      1
     4        10      1
     4        13      1
     5         8      1
     5         9      1
     6         7      1
     7        11      1
     7        12      1
     7        14      1
     7        18      1
     8        10      1
     8        13      1
     9        15      1
     9        16      1
     9        17      1
     9        19      1
     9        20      1
     9        21      1
```

```
      10      14       1
      10      18       1
      11      13       1
      12      13       1
      14      19       1
      14      20       1
      14      21       1
      15      18       1
      16      18       1
      17      18       1

[ angles ]
;    ai      aj      ak    funct      c0          c1

       1       4       5       2   124.27005  600.738720
       1       4       7       2   123.46005  600.822400
       2       1       3       2   115.12005  335.305760
       2       1       4       2   121.15005  408.525760
       3       1       4       2   121.15005  408.525760
       4       5       6       2   110.80005  441.746720
       4       7       8       2   110.80005  441.746720
       4       7       9       2   118.18005  595.048480
       5       4       7       2   113.82005  655.549120
       7       9      10       2   123.46005  600.822400
       7       9      13       2   123.46005  600.822400
       8       7       9       2   110.80005  441.746720
       9      10      11       2   115.09005  406.601120
       9      10      12       2   115.09005  406.601120
       9      13      14       2   123.71005  522.163200
       9      13      18       2   123.71005  522.163200
      10       9      13       2   112.82005  608.353600
      11      10      12       2   115.12005  335.305760
      13      14      15       2   109.79005  414.801760
      13      14      16       2   109.79005  414.801760
      13      14      17       2   109.79005  414.801760
      13      18      19       2   109.79005  414.801760
      13      18      20       2   109.79005  414.801760
      13      18      21       2   109.79005  414.801760
      14      13      18       2   114.51005  529.276000
      15      14      16       2   108.46005  328.360320
      15      14      17       2   108.46005  328.360320
      16      14      17       2   108.46005  328.360320
      19      18      20       2   108.46005  328.360320
      19      18      21       2   108.46005  328.360320
      20      18      21       2   108.46005  328.360320

[ dihedrals ]
```

```
;    ai      aj      ak      al   funct    c0       c1      c2
      1       4       5       6     1    180.0   17.36360   2
      1       4       7       8     1    180.0   17.36360   2
      1       4       7       9     1    180.0   17.36360   2
      2       1       4       5     1    180.0   17.36360   2
      2       1       4       7     1    180.0   17.36360   2
      3       1       4       5     1    180.0   17.36360   2
      3       1       4       7     1    180.0   17.36360   2
      4       7       9      10     1    180.0   17.36360   2
      4       7       9      13     1    180.0   17.36360   2
      5       4       7       8     1    180.0   17.36360   2
      5       4       7       9     1    180.0   17.36360   2
      6       5       4       7     1    180.0   17.36360   2
      7       9      10      11     1    180.0   17.36360   2
      7       9      10      12     1    180.0   17.36360   2
      7       9      13      14     1    180.0   17.36360   2
      7       9      13      18     1    180.0   17.36360   2
      8       7       9      10     1    180.0   17.36360   2
      8       7       9      13     1    180.0   17.36360   2
      9      13      14      15     1    180.0    1.00000   6
      9      13      18      19     1    180.0    1.00000   6
     10       9      13      14     1    180.0   17.36360   2
     11      10       9      13     1    180.0   17.36360   2
     12      10       9      13     1    180.0   17.36360   2
;improper dihedral
      1       4       2       3     2     0.00    4.60240   2
      4       1       7       5     2     0.00    4.60240   2
      9      10       7      13     2     0.00    4.60240   2
     13       9      14      18     2     0.00    4.60240   2
----------------------------------------------------------------------
```

**Coordinates of MET3 (.gro format)**

resid resname  atomtype  resnr      x        y        z

      1MET3      N2       1      2.926    2.715    2.572
      1MET3      H1       2      2.901    2.794    2.516
      1MET3      H3       3      3.024    2.702    2.596
      1MET3      C4       4      2.828    2.635    2.619
      1MET3      N5       5      2.864    2.526    2.691
      1MET3      H6       6      2.960    2.514    2.721
      1MET3      H7       7      2.794    2.477    2.744
      1MET3      N8       8      2.704    2.671    2.591
      1MET3      C9       9      2.596    2.592    2.597
      1MET3     N15      10      2.604    2.457    2.578
      1MET3     H16      11      2.520    2.402    2.565
      1MET3     H17      12      2.686    2.420    2.533

| 1MET3 | N10 | 13 | 2.476 | 2.647 | 2.619 |
| 1MET3 | C18 | 14 | 2.353 | 2.566 | 2.613 |
| 1MET3 | H19 | 15 | 2.351 | 2.490 | 2.692 |
| 1MET3 | H20 | 16 | 2.268 | 2.633 | 2.629 |
| 1MET3 | H21 | 17 | 2.341 | 2.519 | 2.515 |
| 1MET3 | C11 | 18 | 2.460 | 2.789 | 2.652 |
| 1MET3 | H12 | 19 | 2.557 | 2.838 | 2.648 |
| 1MET3 | H13 | 20 | 2.394 | 2.836 | 2.579 |
| 1MET3 | H14 | 21 | 2.418 | 2.799 | 2.752 |

**Topology and Force-Field parameters of MET3 (.itp format)**

```
[ moleculetype ]
; Name             nrexcl
MET3              3

[ atoms ]
;   nr    type  resnr residue  atom   cgnr     charge      mass

     1     na      1    MET3    N2      1   -0.876975    14.0100   ; qtot -0.8770
     2     hn      1    MET3    H1      2    0.430501     1.0080   ; qtot -0.4465
     3     hn      1    MET3    H3      3    0.430501     1.0080   ; qtot -0.0160
     4     cz      1    MET3    C4      4    0.838081    12.0100   ; qtot  0.8221
     5     na      1    MET3    N5      5   -0.876975    14.0100   ; qtot -0.0549
     6     hn      1    MET3    H6      6    0.430501     1.0080   ; qtot  0.3756
     7     hn      1    MET3    H7      7    0.430501     1.0080   ; qtot  0.8061
     8     ne      1    MET3    N8      8   -0.615647    14.0100   ; qtot  0.1905
     9     cz      1    MET3    C9      9    0.613622    12.0100   ; qtot  0.8041
    10     na      1    MET3    N15    10   -0.961600    14.0100   ; qtot -0.1575
    11     hn      1    MET3    H16    11    0.447151     1.0080   ; qtot  0.2897
    12     hn      1    MET3    H17    12    0.447151     1.0080   ; qtot  0.7368
    13     na      1    MET3    N10    13    0.009854    14.0100   ; qtot  0.7467
    14     c3      1    MET3    C18    14   -0.345405    12.0100   ; qtot  0.4013
    15     h1      1    MET3    H19    15    0.157357     1.0080   ; qtot  0.5586
    16     h1      1    MET3    H20    16    0.157357     1.0080   ; qtot  0.7160
    17     h1      1    MET3    H21    17    0.157357     1.0080   ; qtot  0.8733
    18     c3      1    MET3    C11    18   -0.345405    12.0100   ; qtot  0.5279
    19     h1      1    MET3    H12    19    0.157357     1.0080   ; qtot  0.6853
    20     h1      1    MET3    H13    20    0.157357     1.0080   ; qtot  0.8426
    21     h1      1    MET3    H14    21    0.157357     1.0080   ; qtot  1.0000

[ bonds ]
;   ai     aj funct            c0             c1
     1      2     1     0.10100 341749.120000
     1      3     1     0.10100 341749.120000
     1      4     1     0.14010 332544.320000
     4      5     1     0.14010 332544.320000
     4      8     1     0.12920 480406.880000
     5      6     1     0.10100 341749.120000
     5      7     1     0.10100 341749.120000
     8      9     1     0.12920 480406.880000
     9     10     1     0.14010 332544.320000
```

```
     9        13        1    0.14010 332544.320000
    10        11        1    0.10100 341749.120000
    10        12        1    0.10100 341749.120000
    13        14        1    0.14629 274219.360000
    13        18        1    0.14629 274219.360000
    14        15        1    0.10969 276646.080000
    14        16        1    0.10969 276646.080000
    14        17        1    0.10969 276646.080000
    18        19        1    0.10969 276646.080000
    18        20        1    0.10969 276646.080000
    18        21        1    0.10969 276646.080000

[ pairs ]
;    ai       aj funct
      1        6        1
      1        7        1
      1        9        1
      2        5        1
      2        8        1
      3        5        1
      3        8        1
      4       10        1
      4       13        1
      5        9        1
      6        8        1
      7        8        1
      8       11        1
      8       12        1
      8       14        1
      8       18        1
      9       15        1
      9       16        1
      9       17        1
      9       19        1
      9       20        1
      9       21        1
     10       14        1
     10       18        1
     11       13        1
     12       13        1
     14       19        1
     14       20        1
     14       21        1
     15       18        1
     16       18        1
     17       18        1
```

```
[ angles ]
;   ai    aj    ak  funct      c0         c1

     1     4     5     1   109.33005 611.700800
     1     4     8     1   116.14505 633.892736
     2     1     3     1   116.80005 333.632160
     2     1     4     1   119.28005 395.806400
     3     1     4     1   119.28005 395.806400
     4     5     6     1   119.28005 395.806400
     4     5     7     1   119.28005 395.806400
     4     8     9     1   118.18005 595.048480
     5     4     8     1   116.14505 633.892736
     6     5     7     1   116.80005 333.632160
     8     9    10     1   116.14505 633.892736
     8     9    13     1   116.14505 633.892736
     9    10    11     1   119.28005 395.806400
     9    10    12     1   119.28005 395.806400
     9    13    14     1   117.20005 534.296800
     9    13    18     1   117.20005 534.296800
    10     9    13     1   109.33005 611.700800
    11    10    12     1   116.80005 333.632160
    13    14    15     1   108.78005 416.977440
    13    14    16     1   108.78005 416.977440
    13    14    17     1   108.78005 416.977440
    13    18    19     1   108.78005 416.977440
    13    18    20     1   108.78005 416.977440
    13    18    21     1   108.78005 416.977440
    14    13    18     1   125.59005 505.761920
    15    14    16     1   108.46005 328.360320
    15    14    17     1   108.46005 328.360320
    16    14    17     1   108.46005 328.360320
    19    18    20     1   108.46005 328.360320
    19    18    21     1   108.46005 328.360320
    20    18    21     1   108.46005 328.360320

[ dihedrals ]
;   ai    aj    ak    al  funct   c0         c1       c2

     1     4     5     6     1   180.0000   17.36360  2
     1     4     5     7     1   180.0000   17.36360  2
     1     4     8     9     1   160.0000   17.36360  2
     2     1     4     5     1   180.0000   17.36360  2
     2     1     4     8     1   180.0000   17.36360  2
     3     1     4     5     1   180.0000   17.36360  2
     3     1     4     8     1   180.0000   17.36360  2
     4     8     9    10     1   150.0000   17.36360  2
     4     8     9    13     1   160.0000   17.36360  2
```

```
     5        4        8        9     1   160.0000   17.36360    2
     6        5        4        8     1   180.0000   17.36360    2
     7        5        4        8     1   180.0000   17.36360    2
     8        9       10       11     1   180.0000   17.36360    2
     8        9       10       12     1   180.0000   17.36360    2
     8        9       13       14     1   180.0000   17.36360    2
     8        9       13       18     1   180.0000   17.36360    2
     9       13       14       15     1   0.000000    1.00000    2
     9       13       18       19     1   0.000000    1.00000    2
    10        9       13       14     1   180.0000   17.36360    2
    10        9       13       18     1   180.0000   17.36360    2
    11       10        9       13     1   180.0000   17.36360    2
    12       10        9       13     1   180.0000   17.36360    2
    14       13       18       19     1    0.00000    1.00000    2
    15       14       13       18     1    0.00000    1.00000    2
; improper dihedrals
     3        1        4        2     4   180.0000    4.60240    2
     1        5        4        8     4   180.0000    4.60240    2
     4        6        5        7     4   180.0000    4.60240    2
    10       13        9        8     4   180.0000    4.60240    2
     9       11       10       12     4   180.0000    4.60240    2
```
-----------------------------------------------------------------------------------------------------------------

*REFERENCES*

1. Rix, U.; Superti-Furga, G. Target profiling of small molecules by chemical proteomics. *Nat. Chem. Biol.* **2009,** *5*, 616-624.
2. Group, D. P. P. R. Reduction in the incidence of type 2 diabetes with lifestyle intervention or metformin. *N Engl J. Med.* **2002,** *2002*, 393-403.
3. DeVries, J. H.; Bain, S. C.; Rodbard, H. W.; Seufert, J.; D'Alessio, D.; Thomsen, A. B.; Zychma, M.; Rosenstock, J.; Group, L.-D. S. Sequential intensification of metformin treatment in type 2 diabetes with liraglutide followed by randomized addition of basal insulin prompted by A1C targets. *Diabetes care* **2012,** *35*, 1446-1454.
4. Evans, J. M.; Donnelly, L. A.; Emslie-Smith, A. M.; Alessi, D. R.; Morris, A. D. Metformin and reduced risk of cancer in diabetic patients. *Bmj* **2005,** *330*, 1304-1305.
5. Zhou, G.; Myers, R.; Li, Y.; Chen, Y.; Shen, X.; Fenyk-Melody, J.; Wu, M.; Ventre, J.; Doebber, T.; Fujii, N. Role of AMP-activated protein kinase in mechanism of metformin action. *J. Clin. Investig.* **2001,** *108*, 1167.
6. Akinyeke, T.; Matsumura, S.; Wang, X.; Wu, Y.; Schalfer, E. D.; Saxena, A.; Yan, W.; Logan, S. K.; Li, X. Metformin targets c-MYC oncogene to prevent prostate cancer. *Carcinogenesis* **2013,** *34*, 2823-2832.
7. Lonardo, E.; Cioffi, M.; Sancho, P.; Sanchez-Ripoll, Y.; Trabulo, S. M.; Dorado, J.; Balic, A.; Hidalgo, M.; Heeschen, C. Metformin targets the metabolic achilles heel of human pancreatic cancer stem cells. *PloS one* **2013,** *8*, e76518.
8. Shank, J. J.; Yang, K.; Ghannam, J.; Cabrera, L.; Johnston, C. J.; Reynolds, R. K.; Buckanovich, R. J. Metformin targets ovarian cancer stem cells in vitro and in vivo. *Gynecol. Oncol.* **2012,** *127*, 390-397.
9. Garcia, A.; Tisman, G. Metformin, B12, and enhanced breast cancer response to chemotherapy. *J. Clin. Oncol.* **2009,** *28*, e19-e19.
10. Iliopoulos, D.; Hirsch, H. A.; Struhl, K. Metformin decreases the dose of chemotherapy for prolonging tumor remission in mouse xenografts involving multiple cancer cell types. *Cancer research* **2011,** *71*, 3196-3201.
11. Jiralerspong, S.; Palla, S. L.; Giordano, S. H.; Meric-Bernstam, F.; Liedtke, C.; Barnett, C. M.; Hsu, L.; Hung, M.-C.; Hortobagyi, G. N.; Gonzalez-Angulo, A. M. Metformin and pathologic complete responses to neoadjuvant chemotherapy in diabetic patients with breast cancer. *J. Clin. Oncol.* **2009,** *27*, 3297-3302.
12. Hirst, J. A.; Farmer, A. J.; Ali, R.; Roberts, N. W.; Stevens, R. J. Quantifying the effect of metformin treatment and dose on glycemic control. *Diabetes Care* **2012,** *35*, 446-454.
13. Scheurer, M.; Sacher, F.; Brauch, H.-J. Occurrence of the antidiabetic drug metformin in sewage and surface waters in Germany. *J. Environ. Monit.* **2009,** *11*, 1608-1613.
14. Drews, J. Drug discovery: a historical perspective. *Science* **2000,** *287*, 1960-1964.
15. Mukherjee, A.; Sasikala, W. D. Drug-DNA intercalation: From discovery to the molecular mechanism. *Adv. Protein Chem. Struct. Biol* **2013,** *92*, 1-62.
16. Chaires, J. B.; Satyanarayana, S.; Suh, D.; Fokt, I.; Przewloka, T.; Priebe, W. Parsing the free energy of anthracycline antibiotic binding to DNA. *Biochemistry* **1996,** *35*, 2047-2053.
17. Rizzo, V.; Sacchi, N.; Menozzi, M. Kinetic studies of anthracycline-DNA interaction by fluorescence stopped flow confirm a complex association mechanism. *Biochemistry* **1989,** *28*, 274-282.

18. Shahabadi, N.; Heidari, L. Synthesis, characterization and multi-spectroscopic DNA interaction studies of a new platinum complex containing the drug metformin. *Spectrochim. Acta A: Molecular and Biomolecular Spectroscopy* **2014,** *128*, 377-385.
19. Qu, X.; Wan, C.; Becker, H.-C.; Zhong, D.; Zewail, A. H. The anticancer drug–DNA complex: femtosecond primary dynamics for anthracycline antibiotics function. *Proceedings of the National Academy of Sciences* **2001,** *98*, 14212-14217.
20. Remeta, D. P.; Mudd, C. P.; Berger, R. L.; Breslauer, K. J. Thermodynamic characterization of daunomycin-DNA interactions: comparison of complete binding profiles for a series of DNA host duplexes. *Biochemistry* **1993,** *32*, 5064-5073.
21. Carloni, P.; Sprik, M.; Andreoni, W. Key steps of the cis-platin-DNA interaction: density functional theory-based molecular dynamics simulations. *J. Phys. Chem. B* **2000,** *104*, 823-835.
22. Mukherjee, A.; Lavery, R.; Bagchi, B.; Hynes, J. T. On the molecular mechanism of drug intercalation into DNA: a simulation study of the intercalation pathway, free energy, and DNA structural changes. *J. Amer. Chem. Soc.* **2008,** *130*, 9747-9755.
23. Sasikala, W. D.; Mukherjee, A. Molecular mechanism of direct proflavine–DNA intercalation: Evidence for drug-induced minimum base-stacking penalty pathway. *J. Phys. Chem. B* **2012,** *116*, 12208-12212.
24. Wilhelm, M.; Mukherjee, A.; Bouvier, B.; Zakrzewska, K.; Hynes, J. T.; Lavery, R. Multistep drug intercalation: molecular dynamics and free energy studies of the binding of daunomycin to DNA. *J. Amer. Chem. Soc.* **2012,** *134*, 8588-8596.
25. Trieb, M.; Rauch, C.; Wibowo, F. R.; Wellenzohn, B.; Liedl, K. R. Cooperative effects on the formation of intercalation sites. *Nucleic acids research* **2004,** *32*, 4696-4703.
26. Baraldi, P. G.; Bovero, A.; Fruttarolo, F.; Preti, D.; Tabrizi, M. A.; Pavani, M. G.; Romagnoli, R. DNA minor groove binders as potential antitumor and antimicrobial agents. *Medicinal research reviews* **2004,** *24*, 475-528.
27. Rodriguez, R.; Miller, K. M.; Forment, J. V.; Bradshaw, C. R.; Nikan, M.; Britton, S.; Oelschlaegel, T.; Xhemalce, B.; Balasubramanian, S.; Jackson, S. P. Small-molecule–induced DNA damage identifies alternative DNA structures in human genes. *Nat. Chem. Biol.* **2012,** *8*, 301.
28. Childs, S. L.; Chyall, L. J.; Dunlap, J. T.; Coates, D. A.; Stahly, B. C.; Stahly, G. P. A metastable polymorph of metformin hydrochloride: Isolation and characterization using capillary crystallization and thermal microscopy techniques. *Cryst. Growth Des.* **2004,** *4*, 441-449.
29. Shahabadi, N.; Heidari, L. Binding studies of the antidiabetic drug, metformin to calf thymus DNA using multispectroscopic methods. *Spectrochim. Acta A: Molecular and Biomolecular Spectroscopy* **2012,** *97*, 406-410.
30. Frisch, M.; Trucks, G.; Schlegel, H. B.; Scuseria, G.; Robb, M.; Cheeseman, J.; Scalmani, G.; Barone, V.; Mennucci, B.; Petersson, G. Gaussian 09, revision a. 02, gaussian. *Inc., Wallingford, CT* **2009,** *200*.
31. Rappe, A. K.; Goddard III, W. A. Charge equilibration for molecular dynamics simulations. *J. Phys. Chem.* **1991,** *95*, 3358-3363.
32. Wang, J.; Wolf, R. M.; Caldwell, J. W.; Kollman, P. A.; Case, D. A. Development and testing of a general amber force field. *J. Comp. Chem.* **2004,** *25*, 1157-1174.
33. Abraham, M. J.; Murtola, T.; Schulz, R.; Páll, S.; Smith, J. C.; Hess, B.; Lindahl, E. GROMACS: High performance molecular simulations through multi-level parallelism from laptops to supercomputers. *SoftwareX* **2015,** *1*, 19-25.

34. Pearlman, D. A.; Case, D. A.; Caldwell, J. W.; Ross, W. S.; Cheatham, T. E.; DeBolt, S.; Ferguson, D.; Seibel, G.; Kollman, P. AMBER, a package of computer programs for applying molecular mechanics, normal mode analysis, molecular dynamics and free energy calculations to simulate the structural and energetic properties of molecules. *Comp. Phys. Comm.* **1995,** *91*, 1-41.
35. Schrodinger, L. Schrodinger Suite 2012 Induced Fit Docking protocol; Glide version 5.8. *New York, NY: Schrodinger, LLC* **2012**.
36. Pérez, A.; Marchán, I.; Svozil, D.; Sponer, J.; Cheatham, T. E.; Laughton, C. A.; Orozco, M. Refinement of the AMBER force field for nucleic acids: improving the description of α/γ conformers. *Biophys. J.* **2007,** *92*, 3817-3829.
37. Wishart, D. S.; Knox, C.; Guo, A. C.; Eisner, R.; Young, N.; Gautam, B.; Hau, D. D.; Psychogios, N.; Dong, E.; Bouatra, S.; Mandal, R.; Sinelnikov, I.; Xia, J.; Jia, L.; Cruz, J. A.; Lim, E.; Sobsey, C. A.; Shrivastava, S.; Huang, P.; Liu, P.; Fang, L.; Peng, J.; Fradette, R.; Cheng, D.; Tzur, D.; Clements, M.; Lewis, A.; De Souza, A.; Zuniga, A.; Dawe, M.; Xiong, Y.; Clive, D.; Greiner, R.; Nazyrova, A.; Shaykhutdinov, R.; Li, L.; Vogel, H. J.; Forsythe, I. HMDB: a knowledgebase for the human metabolome. *Nucleic Acids Research* **2009,** *37*, D603-D610.
38. Childs, S. L.; Chyall, L. J.; Dunlap, J. T.; Coates, D. A.; Stahly, B. C.; Stahly, G. P. A Metastable Polymorph of Metformin Hydrochloride: Isolation and Characterization Using Capillary Crystallization and Thermal Microscopy Techniques. *Crystal Growth & Design* **2004,** *4*, 441-449.
39. Benmessaoud, I.; Koutchoukali, O.; Bouhelassa, M.; Nouar, A.; Veesler, S. Solvent screening and crystal habit of metformin hydrochloride. *Journal of Crystal Growth* **2016,** *451*, 42-51.
40. Pratt, A. C. The photochemistry of imines. *Chemical Society Reviews* **1977,** *6*, 63-81.
41. Hardy, R. C.; Cottington, R. L. Viscosity of deuterium oxide and water in the range 5 to 125 C. *J Res Natl Bureau Standards* **1949,** *42*, 573-578.
42. Socrates, G. *Infrared and Raman Characteristic Group Frequencies*. John Wiley and Sons, Ltd: 2001.
43. Luzar, A.; Chandler, D. Hydrogen-bond kinetics in liquid water. *Nature* **1996,** *379*, 55-57.
44. Bagchi, B. *Molecular relaxation in liquids*. OUP USA: 2012.
45. Maroncelli, M.; Fleming, G. R. Picosecond solvation dynamics of coumarin 153: the importance of molecular aspects of solvation. *J. Chem. Phys* **1987,** *86*, 6221-6239.
46. Bagchi, B.; Chandra, A. Ultrafast solvation dynamics: Molecular explanation of computer simulation results in a simple dipolar solvent. *J. Chem. Phys* **1992,** *97*, 5126-5131.
47. Bagchi, B.; Jana, B. Solvation dynamics in dipolar liquids. *Chem. Soc. Rev.* **2010,** *39*, 1936-1954.
48. Bhattacharyya, K. Solvation dynamics and proton transfer in supramolecular assemblies. *Acc. Chem. Res.* **2003,** *36*, 95-101.
49. Bagchi, B. *Water in Biological and Chemical Processes: From Structure and Dynamics to Function*. Cambridge University Press: 2013.
50. Maroncelli, M.; Fleming, G. R. Computer simulation of the dynamics of aqueous solvation. *J. Chem. Phys* **1988,** *89*, 5044-5069.
51. Mondal, S.; Mukherjee, S.; Bagchi, B. Origin of diverse time scales in the protein hydration layer solvation dynamics: A simulation study. *J. Chem. Phys* **2017,** *147*, 154901.

52. Barducci, A.; Bonomi, M.; Parrinello, M. Metadynamics. *Wiley Interdiscip. Rev. Comput. Mol. Sci* **2011,** *1*, 826-843.
53. Barducci, A.; Bussi, G.; Parrinello, M. Well-tempered metadynamics: a smoothly converging and tunable free-energy method. *Phys. Rev. Lett.* **2008,** *100*, 020603.
54. Bonomi, M.; Branduardi, D.; Bussi, G.; Camilloni, C.; Provasi, D.; Raiteri, P.; Donadio, D.; Marinelli, F.; Pietrucci, F.; Broglia, R. A. PLUMED: A portable plugin for free-energy calculations with molecular dynamics. *Comp. Phys. Comm.* **2009,** *180*, 1961-1972.
55. Tribello, G. A.; Bonomi, M.; Branduardi, D.; Camilloni, C.; Bussi, G. PLUMED 2: New feathers for an old bird. *Comp. Phys. Comm.* **2014,** *185*, 604-613.
56. Chaires, J. B. Drug—DNA interactions. *Curr. Opin. Struct. Biol.* **1998,** *8*, 314-320.
57. Fresch, B.; Remacle, F. Atomistic account of structural and dynamical changes induced by small binders in the double helix of a short DNA. *Physical Chemistry Chemical Physics* **2014,** *16*, 14070-14082.
58. Haq, I. Thermodynamics of drug–DNA interactions. *Archives of biochemistry and biophysics* **2002,** *403*, 1-15.
59. Adam, G.; Gibbs, J. H. On the temperature dependence of cooperative relaxation properties in glass-forming liquids. *J. Chem. Phys* **1965,** *43*, 139-146.
60. Rosenfeld, Y. Relation between the transport coefficients and the internal entropy of simple systems. *Physical Review A* **1977,** *15*, 2545.
61. Mukherjee, A. Entropy balance in the intercalation process of an anti-cancer drug daunomycin. *The Journal of Physical Chemistry Letters* **2011,** *2*, 3021-3026.
62. Humphrey, W.; Dalke, A.; Schulten, K. VMD: visual molecular dynamics. *J. Mol. Graph.* **1996,** *14*, 33-38.
63. Cantor, C. R.; Schimmel, P. R. *Biophysical Chemistry - Techniques for the Study of Biological Structure and Function*. W. H. Freeman and Company: New York, 2001; Vol. II, p 376-379.
64. Strekowski, L.; Wilson, B. Noncovalent interactions with DNA: an overview. *Mutation Research/Fundamental and Molecular Mechanisms of Mutagenesis* **2007,** *623*, 3-13.
65. Shahabadi, N.; Maghsudi, M. Multi-spectroscopic and molecular modeling studies on the interaction of antihypertensive drug; methyldopa with calf thymus DNA. *Mol. BioSyst.* **2014,** *10*, 338-347.
66. Garbett, N. C.; Ragazzon, P. A.; Chaires, J. B. Circular dichroism to determine binding mode and affinity of ligand-DNA interactions. *Nat. Protocols* **2007,** *2*, 3166-3172.
67. Gray, D. M.; Ratliff, R. L.; Vaughan, M. R. [19] Circular dichroism spectroscopy of DNA. In *Methods in Enzymology*, Academic Press: 1992; Vol. 211, pp 389-406.
68. Long, E. C.; Barton, J. K. On demonstrating DNA intercalation. *Accounts of Chemical Research* **1990,** *23*, 271-273.
69. Allenmark, S. Induced circular dichroism by chiral molecular interaction. *Chirality* **2003,** *15*, 409-422.
70. Chen, X.; Liu, M. Induced chirality of binary aggregates of oppositely charged water-soluble porphyrins on DNA matrix. *J Inorg. Biochem* **2003,** *94*, 106-113.
71. Abraham, M.; Van Der Spoel, D.; Lindahl, E.; Hess, B. The GROMACS development team GROMACS user manual version 5.0. 4. Accessed: 2014.